\documentclass[manuscript,screen]{acmart}
\AtBeginDocument{%
  }

\setcopyright{acmlicensed}
\copyrightyear{2026}
\acmYear{2026}
\acmDOI{XXXXXXX.XXXXXXX}
\acmConference[CHI '26]{ACM CHI conference on Human Factors in Computing Systems}{April 13--17,
  2026}{Barcelona, Spain}

\acmISBN{978-1-4503-XXXX-X/2018/06}

\usepackage{graphicx}
\usepackage{subcaption}
\usepackage{wrapfig}
\usepackage{multirow}
\graphicspath{{figures/}}

\usepackage{xcolor, colortbl}

\usepackage{cleveref}

\begin{document}

\title{Human-AI Narrative Synthesis to Foster Shared Understanding in Civic Decision-Making}

\author{Cassandra Overney}
\authornote{Corresponding author}
\email{coverney@mit.edu}
\affiliation{%
  \institution{Massachusetts Institute of Technology}
  \city{Cambridge}
  \state{MA}
  \country{USA}
}
\affiliation{%
  \institution{Northeastern University}
  \city{Boston}
  \state{MA}
  \country{USA}
}

\author{Hang Jiang}
\email{hjian42@mit.edu}
\affiliation{%
  \institution{Massachusetts Institute of Technology}
  \city{Cambridge}
  \state{MA}
  \country{USA}
}
\affiliation{%
  \institution{Northeastern University}
  \city{Boston}
  \state{MA}
  \country{USA}
}

\author{Urooj Haider}
\email{lnu.ur@northeastern.edu}
\affiliation{%
  \institution{Northeastern University}
  \city{Boston}
  \state{MA}
  \country{USA}
}

\author{Cassandra Moe}
\email{moe.c@northeastern.edu}
\affiliation{%
  \institution{Northeastern University}
  \city{Boston}
  \state{MA}
  \country{USA}
}

\author{Jasmine Mangat}
\email{j.mangat@northeastern.edu}
\affiliation{%
  \institution{Northeastern University}
  \city{Boston}
  \state{MA}
  \country{USA}
}

\author{Frank Pantano}
\email{fnpantano@wsfcs.k12.nc.us}
\affiliation{%
  \institution{Winston-Salem/Forsyth County Schools}
  \city{Winston-Salem}
  \state{NC}
  \country{USA}
}

\author{Effie G. McMillian}
\email{egmcmillian@wsfcs.k12.nc.us}
\affiliation{%
  \institution{Winston-Salem/Forsyth County Schools}
  \city{Winston-Salem}
  \state{NC}
  \country{USA}
}

\author{Paul Riggins}
\authornote{Co–last author}
\email{p.riggins@northeastern.edu}
\affiliation{%
  \institution{Northeastern University}
  \city{Boston}
  \state{MA}
  \country{USA}
}

\author{Nabeel Gillani}
\authornotemark[2]
\email{n.gillani@northeastern.edu}
\affiliation{%
  \institution{Northeastern University}
  \city{Boston}
  \state{MA}
  \country{USA}
}

\renewcommand{\shortauthors}{Overney et al.}

\begin{abstract}
Community engagement processes in representative political contexts, like school districts, generate massive volumes of feedback that overwhelm traditional synthesis methods, creating barriers to shared understanding not only between civic leaders and constituents but also among community members.
To address these barriers, we developed StoryBuilder, a human-AI collaborative pipeline that transforms community input into accessible first-person narratives.
Using 2,480 community responses from an ongoing school rezoning process, we generated 124 composite stories and deployed them through a mobile-friendly StorySharer interface. 
Our mixed-methods evaluation combined a four-month field deployment, user studies with 21 community members, and a controlled experiment examining how narrative composition affects participant reactions.
Field results demonstrate that narratives helped community members relate across diverse perspectives.
In the experiment, experience-grounded narratives generated greater respect and trust than opinion-heavy narratives.
We contribute a human-AI narrative synthesis system and insights on its varied acceptance and effectiveness in a real-world civic context.
\end{abstract}


\begin{teaserfigure}
  \includegraphics[width=\textwidth]{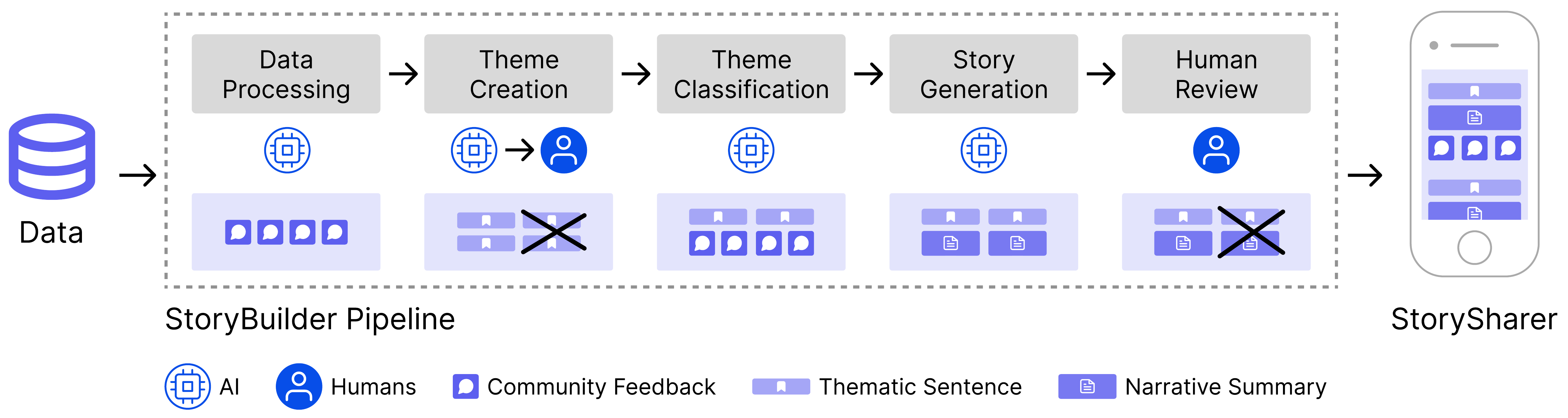}
  \caption{StoryBuilder Pipeline and StorySharer Interface Overview. The system transforms community voice data (N=2,480 feedback segments) through a five-stage human-AI collaborative pipeline, organized by topic and stakeholder type: (1) data processing with scene/theme extraction, (2) AI theme creation with human editing, (3) theme classification of data, (4) AI story generation by theme, and (5) human review and editing. The resulting stories are deployed through the StorySharer interface.}
  \Description{
    The figure presents the StoryBuilder pipeline and StorySharer interface. On the left, community feedback data (2,480 feedback segments) enters a five-stage human-AI collaborative pipeline. The stages are shown as sequential boxes: (1) Data Processing, which extracts scenes and themes; (2) Theme Creation, where the AI suggests themes and humans edit them; (3) Theme Classification, which assigns data segments to themes; (4) Story Generation, where the AI drafts stories based on themes; and (5) Human Review, where humans edit and finalize the stories. The final stories flow into the StorySharer interface on the right framed by a mobile device, where they are displayed as thematic sentence- and narrative-style summaries. Icons and labels indicate where AI and humans are involved throughout the process, along with the flow from community feedback to thematic sentences to narrative summaries.
  }
  \label{fig:teaser}
\end{teaserfigure}


\maketitle

\section{Introduction}
Community engagement processes are critical for democratic governance across contexts from school districts to urban planning, yet these processes generate massive volumes of qualitative feedback through thousands of individual responses that overwhelm traditional synthesis methods~\cite{mahyarCivicDataDeluge2019}. 
While civic leaders invest significant resources in gathering community voices, the dual challenges of synthesizing feedback at scale~\cite{jasimCommunityPulseFacilitatingCommunity2021} and sharing insights in accessible formats~\cite{overney2025boundarease, chandler2015listening} remain persistent barriers to promoting shared understanding.
This synthesis-sharing challenge is particularly acute in representative political contexts, such as school districts and municipal governments, where civic leaders must synthesize community input before sharing results with constituents, unlike more participatory referendum or assembly contexts~\cite{nelimarkka2019review}.
This raises a critical research question: \textit{How can we effectively summarize large-scale qualitative feedback data to share with a community in a way that fosters connection and understanding across diverse perspectives?}

The HCI community has developed various community engagement platforms for collecting, analyzing, and sharing community input~\cite{gunUrbanDesignEmpowerment2020, saldivarCivicTechnologySocial2019}. 
Recent advances in AI, particularly large language models (LLMs), offer promising synthesis capabilities but raise critical concerns about accuracy, transparency, and bias~\cite{pengPathwayUrbanPlanning2023}, highlighting the need for human-AI collaborative solutions.
Most existing human-AI systems focus on either internal sensemaking support for officials (e.g., \cite{overneySenseMateAccessibleBeginnerFriendly2024}) or external sharing with communities (e.g., \cite{kripleanSupportingReflectivePublic2012}). This leaves a gap in integrated approaches that promote shared understanding between decision-makers and constituents. 
Limited exceptions either focus on participatory contexts that emphasize direct deliberation~\cite{smallPolisScalingDeliberation2021} or create civic infrastructure that primarily surfaces raw community input without synthesis support~\cite{hughes2025voice}---yet these approaches typically still do not engage communities in making meaning out of others' contributions and subsequently understanding where their own contributions fit in. This is essential for enhancing awareness and empathy in settings where there are divergent views about specific issues. 
Narrative approaches in particular can hold promise for promoting understanding and connection across diverse viewpoints more effectively than opinions~\cite{dahlstrom2014using, kubinPersonalExperiencesBridge2021}.
Building on these insights, in this project, we leverage AI-supported qualitative analysis methods and narrative frameworks to support scalable sensemaking in representative political contexts, while preserving authenticity and accessibility through storytelling.

We developed a human-AI collaborative StoryBuilder pipeline that employs LLMs to transform qualitative community input into first-person ``composite'' stories comprised of multiple community members' feedback, maintaining citations to original stakeholder input. These stories are closely evaluated by humans before deployment.
Our pipeline seeks to simultaneously support synthesis and sharing of community voices in an accessible format.
%
We implemented our pipeline using a real-world dataset of 2,480 community stories and experiences from an ongoing and highly contentious school rezoning process, conducted in a large school district ($\sim$50,0000 students) that has not made substantial student zoning assignment changes in over a generation. Using this input, we generated 124 composite narrative summaries and deployed them back to the community through a mobile-friendly StorySharer interface. 
Our evaluation employed a mixed-methods approach combining a four-month field deployment (2,183 sessions), user studies with 21 community members, and a pre-registered, controlled experiment examining how narrative structure variations affect interpersonal measures of connection, understanding, respect, and trust.

Usage patterns in the field deployment showed high mobile adoption (51\%) and sustained engagement (6.2 minutes average), though voice exploration remained limited with only 12\% of users interacting with citations---even though citations were valued by many user study participants.
Our analysis of platform usage patterns and accompanying user study interviews revealed that narratives
helped community members relate
across diverse perspectives, even when stories did not align with readers' experiences. 
However, we observed mixed results regarding feelings of connection to people with different experiences.
While some user study participants valued cross-stakeholder perspectives, feedback analytics on the field deployment
suggest that users showed lower connections ratings to stakeholder groups outside their own.
Our controlled experiment enabled us to more systematically understand the impact of different types of stories on several key outcomes. In particular, we found that experience-heavy narratives fostered greater respect and trust across differences than opinion-heavy stories, confirming findings from prior work~\cite{kubinPersonalExperiencesBridge2021, hagmann2024personal}. 
These effects did not extend to changes in participants' positions on diversity in education,
suggesting that while narrative framing can build interpersonal foundations for constructive engagement, it may not directly alter policy preferences.
Experiment participants also expressed some notable critique of AI use for writing narrative summaries, but community members in the actual context of field deployment expressed broad acceptance, given the need to summarize so much feedback from their community.
%
%
This work contributes to HCI and civic technology in three ways: 
\begin{enumerate}
    \item Human-AI Collaborative Pipeline: A scalable narrative synthesis system that maintains authenticity through comprehensive source attribution and human oversight, demonstrating how LLMs can transform qualitative community feedback into accessible first-person stories
    \item Field Deployment Insights: Real-world findings on how narrative strategies might impact community engagement in contentious and high-stakes, polarized policymaking settings, including detailed usage analytics and qualitative community insights
    \item Experimental Evidence: Controlled study findings on how different narrative compositions affect understanding and connection
\end{enumerate}

\subsection{Positionality Statement}

Our research team brings together academic researchers from human-computer interaction, computational social science, and education with practicing school administrators who have direct responsibility for rezoning policy processes. 
Two of our co-authors work within the school district where we deployed our system, serving in leadership roles during the rezoning process studied, while five additional co-authors were hired by the district as external consultants who assisted with community engagement and generated potential boundary scenarios, providing deep contextual knowledge of the district's specific challenges. 
The remaining two co-authors maintained greater distance from the field work, contributing research-oriented perspectives to balance the team's immersion in the field. 
We acknowledge that our team's dual role as researchers and practitioners influenced our system development, particularly accelerating our story review timeline to meet district commitments for community feedback reporting.
We attempted to minimize potential bias by conducting user studies and a controlled experiment to triangulate our field deployment findings. These activities also helped ensure that research and academic freedom were not subsumed by the team's contractual obligations as service providers in the district. We believe the different backgrounds represented across the project team helped ensure that research and practice-oriented considerations were carefully balanced and duly attended to throughout the study.

\section{Background}

\subsection{Digital Tools for Community Engagement}

Public participation is crucial for integrating community perspectives into local policy decisions~\cite{corbettProblemCommunityEngagement2018}. 
The HCI community has developed systems to facilitate community engagement across three core activities: collecting, processing, and sharing community input~\cite{gunUrbanDesignEmpowerment2020, saldivarCivicTechnologySocial2019}.
Feedback collection has been supported through systems for town halls~\cite{jasimCommunityClickCapturingReporting2021} and collaborative deliberation~\cite{johnsonCommunityConversationalSupporting2017, mahyarUDCoSpacesTableCentred2016}. 
%
Once collected, feedback requires systematic analysis. 
Multiple platforms support this through qualitative visualization tools~\cite{jasimCommunityPulseFacilitatingCommunity2021, beefermanFeedbackMap2023} and sensemaking interfaces~\cite{overneySenseMateAccessibleBeginnerFriendly2024, hughes2025voice}. 
However, civic leaders face scalability challenges when processing extensive qualitative datasets~\cite{reynanteFrameworkOpenCivic2021, mahyarCivicDataDeluge2019}.
To complete the engagement loop, some platforms return processed insights to communities through interactive reports and dashboard~\cite{mayDesignCivicTechnology2018, maskellSpokespeopleExploringRoutes2018, hughes2025voice}. 
These systems implement various strategies to facilitate exposure to different perspectives~\cite{kripleanSupportingReflectivePublic2012, nelimarkkaComparingThreeOnline2014}, though analysis results often appear in inaccessible formats that undermine meaningful engagement~\cite{overney2025boundarease}.

Current limitations create opportunities for AI-assisted approaches, particularly through large language models~\cite{overney2025designing}. 
However, deploying AI in civic contexts raises concerns about accuracy~\cite{huang2023survey}, transparency~\cite{smallOpportunitiesRisksLLMs2023}, and bias~\cite{pengPathwayUrbanPlanning2023}. 
While existing systems have explored AI support for specific functions~\cite{overney2025coalesce} and participatory contexts~\cite{leeWeBuildAIParticipatoryFramework2019, smallOpportunitiesRisksLLMs2023}, there remains a gap in supporting representative political contexts where civic leaders manage sensemaking before sharing results with constituents~\cite{nelimarkka2019review}.
Our work contributes to this space by developing an AI-mediated pipeline that simultaneously supports synthesis and sharing through narrative sensemaking of community feedback, maintaining human oversight while leveraging computational scale to bridge internal analysis and external community engagement.

\subsection{Narrative and Storytelling in HCI}

Narrative has long been used in HCI to foster engagement, empathy, and comprehension. Designers employ stories such as personas, scenarios, and design fictions to ground technology in lived experience and anticipate futures \cite{carroll1997scenario,tanenbaum2014design}. Narratives also improve communication barriers compared to expository formats, from science communication to legal education \cite{dahlstrom2014using,jiang-etal-2024-leveraging}. NLP researchers have studied how to generate narratives from plot-based methods like Plan-and-Write and GraphPlan \cite{yao2019plan,chen2021graphplan} to fluent, emotionally rich narratives enabled by large language models \cite{xie2023next}. Applications include children’s learning \cite{valentini2023automatic}, accessible legal narratives \cite{jiang-etal-2024-leveraging}, and personality-reflective stories \cite{jiang-etal-2024-personallm}. Such work highlights the educational, social, and emotional potential of generative AI.

A central consideration in narrative design is perspective. While many stories adopt a third-person voice, narratology shows that first- or second-person narration increases self-involvement and emotional resonance \cite{mildorf2012second,mildorf2016reconsidering}. In addition, McAdams’ framework distinguishes between \textit{scenes} (experiences), \textit{themes} (interpretations), and \textit{tone} (emotional tenor) \cite{mcadams2001psychology}. Experience-heavy stories emphasize scenes, while opinion-heavy stories emphasize themes, a useful lens for structuring LLM-generated narratives. Storytelling also shapes data-rich contexts: narrative visualizations blend author-driven explanation with reader exploration to guide insight \cite{segel2010narrative,hullman2011visualization}. Authenticity further influences trust: first-person testimonials feel more credible than synthetic accounts, while perceived artificiality can reduce engagement \cite{lambert2013digital}. These concerns are acute for AI-authored content, raising questions about authorship and transparency.

In sum, research across HCI, narratology, and NLP shows that narrative enhances comprehension, attention, and emotional connection. Choices around perspective, data integration, and authenticity shape user experience. Building on McAdams’ scenes–themes–tone framework, we present LLM-generated stories to examine both their engagement benefits and the tensions users perceive around authenticity and relatability.

\subsection{AI-Supported Qualitative Data Analysis}

To generate the narratives grounded in community voice, we developed an AI-supported thematic analysis approach, building off of related work in this space. Traditional qualitative analysis methods, like thematic analysis and grounded theory~\cite{braun2006using, glaser1968discovery, charmaz2006constructing}, produce rich and rigorous understanding but are difficult to scale to large datasets~\cite{chandrasekar2024making}. To handle large datasets, researchers have developed a variety of computational methods, like topic modeling, clustering, sentiment analysis, and varied machine learning approaches~\cite{lennon2021developing, abram2020methods, drouhard2017aeonium, grandeit2020using}. Generating plain language thematic descriptions from these methods, however, still requires manual researcher intervention (e.g., \cite{baumer2017comparing}), which can be difficult to scale when the number of themes is large.

More recently, LLMs have arisen as a potent computational tool for natural language processing. LLM-based tools show promise at supporting many aspects of qualitative data analysis and sensemaking~\cite{jiang2024bridging, ye2025scholarmate, kabbara2025ai, hitch2024artificial, morgan2023exploring, gaocollabcoder, gebreegziabher2023patat, zhang2023qualigpt}, especially at scale, but also raise new concerns around accuracy, data privacy, and bias~\cite{arguedas2023automating, chandrasekar2024making}. Their popularity also redoubles concerns raised with earlier computational methods, around whether computational methods diminish deep reading and scholarly engagement with the source data. As with pre-LLM computational methods~\cite{jiang2021supporting, grimmer2013text, rieder2012digital, carlsen2022computational, nelson2020computational}, current best practices seem to recommend keeping human researchers closely in the loop to guide, correct, and validate computational results.

\section{Narrative Synthesis System}

Building on recommendations for human-in-the-loop analysis approaches, we developed a human-AI collaborative system to synthesize large-scale community feedback into narrative summaries. 
Figure~\ref{fig:teaser} provides an overview of this process, which generated 124 community stories from 2,480 initial feedback segments collected during an active school rezoning process. 
Below, we describe each component: the community feedback dataset, the StoryBuilder technical pipeline, and the StorySharer interface.

\subsection{Data}


The dataset for this study came from a redistricting project with a large ($\sim$50,000 students) school district.
The school district underwent a two-year planning project to update its residential school zone boundaries, which had not been comprehensively reviewed since the 1990s. (At the time of this writing, the project is still in its final stages.) The prospect of changing these boundaries is highly controversial in the community. A core part of this project involved intensive community engagement to understand community thoughts and feelings about the school district in general and the possibility of redistricting in particular. The data from that engagement forms the basis for the study. (See \Cref{app:data} for more details.)

Community engagement data had two primary components: (1) $\sim$8,400 responses from a district-wide online survey and (2) $\sim$170 hours of audio from a series of in-person and virtual facilitated feedback sessions. After processing, the combined survey and facilitated session data yielded 15,104 distinct quotes ($\sim$809k words) from write-in responses and segmented recording transcripts. These quotes were coded by topic, like ``Transportation'' or ``Student Well-Being'', in an AI-assisted qualitative analysis process separate from this project. Quotes were also coded by feedback type, like ``opinion'', ``suggestion'', or ``personal experience''. We chose to use the ``story'' and ``personal experience'' quotes as the most appropriate source material to generate the composite stories in this study, totaling 2,480 feedback quotes ($\sim$358k words).

District leaders and community members wanted to know, ``What is the community saying?'' We wanted a solution that preserved the voices of the community and reflected the large quantity, as well as rich variety, of views they shared, while respecting district timelines for closing the community engagement feedback loop. Our StoryBuilder pipeline creates a way to do this efficiently and effectively.

\subsection{StoryBuilder Pipeline}
Our system transforms raw community feedback into aggregate narratives through a five-stage pipeline combining large language model (LLM) automation with human review.
Each story is created from community quotes around a common theme, organized by topic and stakeholder type for dissemination.
The technical pipeline leverages DSPy~\cite{khattab2023dspy} for structured LLM interactions, enabling systematic prompt evaluation.
Total processing costs were around \$250.
The supplementary materials contains the code and LLM prompts, along with details regarding model performance throughout the pipeline.

\subsubsection{Data Processing}

We preprocessed community responses by extracting 2,480 individual stories and experiences as ``building blocks'' for narrative generation. 
56\% originated from survey responses, with the remainder from listening session transcripts. 
Using GPT-4o-mini, we applied the McAdams Life Story Framework~\cite{mcadams2001psychology} to decompose each building block into scenes (concrete events and experiences) and themes (underlying interpretations and values). 
This decomposition provides structured narrative components to preserve authentic details from the original feedback. 

\subsubsection{Theme Creation}
\label{sec:theme creation}

We aimed to minimize the number of unique themes to reduce information overload in dissemination, while still maintaining full coverage of diverse community perspectives. Building on previous thematic analysis that had been done on the dataset, we initially organized the data into 18 topics (Diversity, Transportation, Student Well-Being, etc.). We used LLMs (both \texttt{Claude 3.5 Sonnet} and \texttt{GPT-4o}) to extract lists of detailed themes within each topic and stakeholder type. The combined initial lists from both models totaled 1,048 themes from the 2,480 stories and experiences with many redundancies.
The author most familiar with the community data used a mix of \texttt{GPT-4o} and manual review to identify and reduce overlaps between topics and themes.
Ultimately, we consolidated down to 190 themes divided across 9 topics and the three stakeholder groups.
These themes would later be revisited and edited again to reduce the reading level and improve theme-story alignment for public dissemination (see \Cref{sec:human review}).

Our LLM-based implementation additionally categorized themes into four types following frameworks developed in prior work~\cite{jasimCommunityClickCapturingReporting2021, jasimCommunityPulseFacilitatingCommunity2021}: pluses (positive current experiences), deltas (negative current experiences), hopes (future aspirations), and concerns (future worries). After theme generation, we found the delta/concern distinction to be too subtle in practice, and ultimately consolidated to just pluses, hopes, concerns.

\subsubsection{Theme Classification}

Once themes under each topic-stakeholder combination were reviewed and finalized, we mapped building blocks to themes by implementing a multi-pass classification system using \texttt{GPT-4o-mini}. 
A theme classification prompts was applied to each building block three times with slight temperature variations.
We computed consensus through set intersection across iterations, requiring agreement across all passes for final theme assignment. 
This approach reduces noise while increasing precision in theme classification.

\subsubsection{Story Generation}

We used \texttt{Claude 3.5 Sonnet} to synthesize building blocks into first-person narratives, told as aggregate stories with all scenes and views attributed to a single first-person narrator. 
Our prompt engineering approach balances concrete scenes from source material with interpretive themes, following the McAdams framework \cite{mcadams2001psychology}.
For each of the 190 themes identified earlier, the pipeline identifies 3-5 quotes classified under that theme that could create a cohesive aggregate narrative.
It then knits together scenes from those quotes (extracted earlier during data processing) with the theme that these quotes share in common to create a balance of narrative details and interpretive views.
Stories maintain source attribution through citations with unique identifiers that map to original community feedback. 
In addition, the prompt requests school name redaction and a fifth-grade reading level for increased accessibility. 
Each story went through a round of automated validation for: (1) minimum three unique citations that are accurately cited, (2) thematic relevance, (3) narrative coherence, (4) stakeholder authenticity, and (5) reading accessibility.
The supplementary materials provides more details on the automated validation process.

\subsubsection{Human Review}
\label{sec:human review}

Stories, along with theme information and results from the automated validation, were prepared for manual review by distributing the nine topics among the authors.
For each topic, a reviewer read through each story and theme to decide which stories should be removed, edited, or preserved.
Reviewers were asked to consider the same dimensions of relevance, coherence, readability, believability, and citation accuracy as the automated validation process. 
In addition, reviewers considered potential duplication between stories within a topic. 
This process happened in three passes: First, reviewers noted and removed any redundant and low quality stories. Second, reviewers made minor edits to improve remaining stories. Third, reviewers checked citations for accuracy on a non-random subset of stories (e.g., stories with a surprising claim or narrative detail). 
Of the 190 AI-generated stories, 66 were dropped; 44 were edited; and 80 were left as is.

Reviewers also revised the final set of themes, expanding it to 301 themes covering all topics and stakeholder groups. 124 of the themes corresponded to the final stories and would serve as story titles; some of these were edited to improve title-story alignment. The other 177 themes captured additional issues that were prominent in community feedback, but for which we did not have enough building blocks to construct high quality stories. In most cases where a story was dropped, for instance, the theme was preserved. We also performed the theme extraction process from \Cref{sec:theme creation} on the rest of the dataset (opinions, etc.), outside the 2,480 quotes we used as story-building material, which yielded several additional themes. Finally, the two authors most familiar with the community data reviewed the themes and added themes that were conspicuously missing. All themes were reduced to an 8th grade reading level. 

A key consideration in this final review process was maintaining a balance of perspectives in each topic and stakeholder group, because of the high community tensions and disagreements around aspects of the redistricting project. We aimed to faithfully represent both supportive and critical perspectives that appeared in each topic.

\subsubsection{Technical Evaluation of AI-Generated Stories} 

Six authors conducted a human evaluation of 20 LLM-generated stories across six binary (Yes/No) dimensions: (1) relevance to the theme, (2) appropriate citation, (3) correct citation, (4) coherence, (5) readability, and (6) believability of the stakeholder voice. Three annotators independently assessed each story. We report exact match agreement, calculated as the pairwise average among the three annotators\footnote{We did not use Cohen’s kappa because the limited variability in binary responses makes it less suitable.}. Agreement scores were generally high across dimensions, with the strongest reliability on readability (0.98), believability (0.93), and appropriate citation (0.85), while coherence (0.80) and relevance (0.82) were slightly lower, and correct citation showed the lowest agreement (0.68). Averaged across stories, the ratings indicate that the generated outputs were highly readable ($0.99 \pm 0.09$), believable ($0.97 \pm 0.18$), and largely relevant ($0.88 \pm 0.33$) and coherent ($0.84 \pm 0.37$). These scores are reported as mean $\pm$ standard deviation across annotators. Citations were mostly judged appropriate ($0.92 \pm 0.28$), though the accuracy of citation attribution was more variable ($0.55 \pm 0.50$). This suggests frequent errors or mismatches between source material and quoted content. Such variability reflects a broader challenge of ``citation hallucination'' in retrieval-augmented generation (RAG), where models fabricate or misattribute references despite fluent text \cite{qian2024capacity,song2024rag, huang2024training,sun2024redeep,schreieder2025attribution}. Taken together, these results indicate that while LLMs reliably produce fluent and plausible stakeholder stories, citation accuracy remains a significant weakness requiring more systematic mitigation.

Citation analysis across the 124 generated stories reveals comprehensive source attribution with 563 total citations referencing 401 unique community responses, indicating selective reuse of 162 sources (29\%) across multiple stories.
Stories on average included a comparable number of 4.54 (SD = 1.02) citations. 
Citations averaged 237 (SD = 210) words, reflecting substantial excerpts from original feedback rather than brief quotes. 
Source distribution favored listening session transcripts over survey responses (62\% vs. 38\%), while stakeholder representation was roughly balanced across students (33\%), parents (32\%), staff (19\%), and dual parent/staff roles (17\%). 

\subsection{StorySharer Interface}

\begin{figure}[t]
    \centering
    \includegraphics[width=\linewidth]{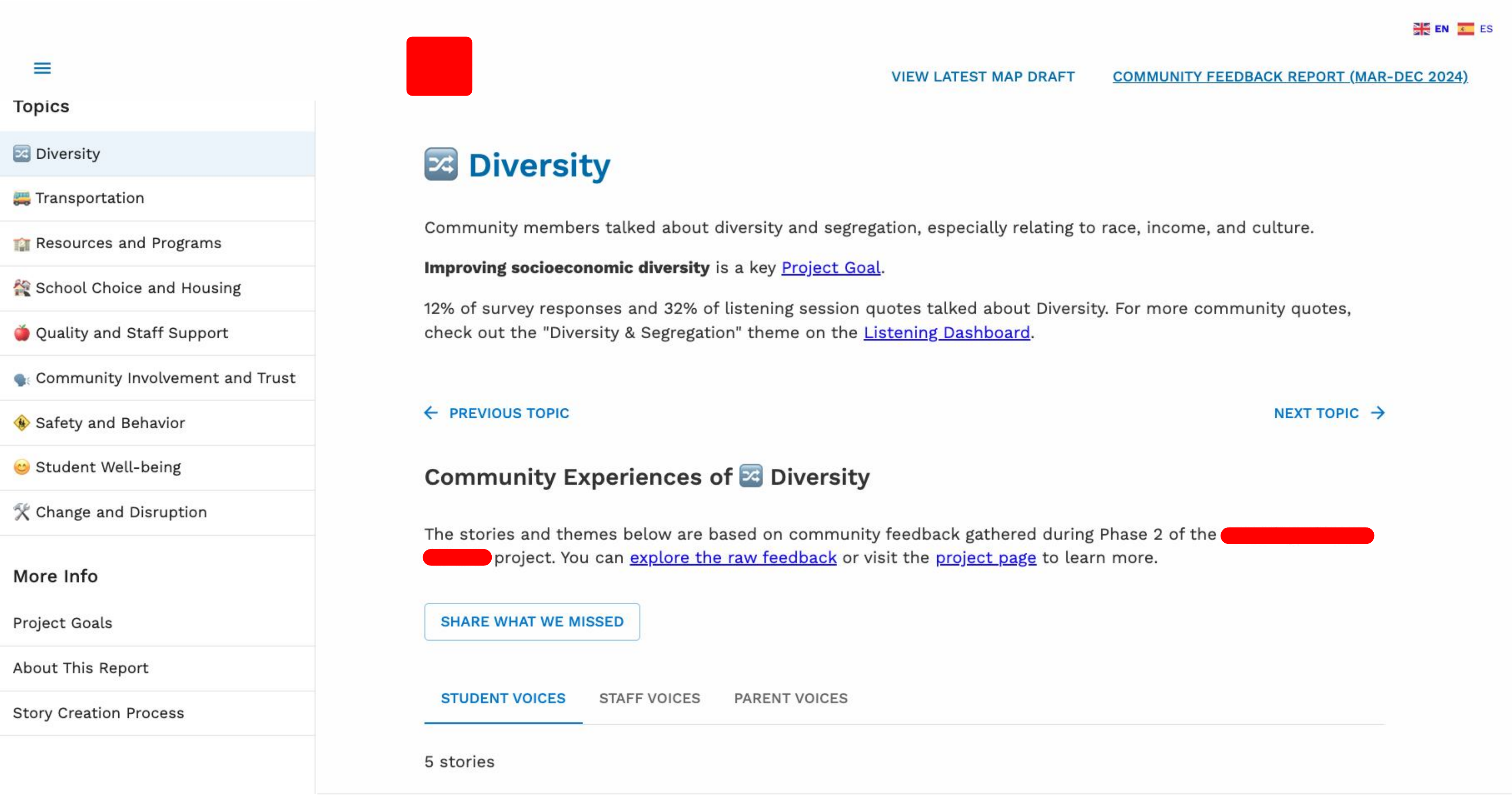}
    \caption{Overall interface layout enabling systematic exploration of 124 community stories organized by topic and stakeholder perspective.}
    \label{fig:platform}
    \Description{
        Image displays a screenshot of a webpage. On the left is a navigation panel with two main sections: Topics and More Info. The Topics section displays a series of nine pages in the following order: Diversity, Transportation, Resources and Programs, School Choice and Housing, Quality and Staff Support, Community Involvement and Trust, Safety and Behavior, Student Well-being, Change and Disruption. The More Info section has the following pages: Project Goals, About This Report, and Story Creation Process. Currently the Diversity page is selected and opened on the right.

        On the right, starting at the top, there is an option to toggle between viewing the page content in English or Spanish. Under that, a description of the diversity topic is provided as follows: Community members talked about diversity and segregation, especially relating to race, income, and culture. Improving socioeconomic diversity is a key Project Goal (with hyperlink to Project Goals page). 12\% of survey responses and 32\% of listening session quotes talked about Diversity. For more community quotes, check out the "Diversity \& Segregation" theme on the Listening Dashboard (with hyperlink to a separate listening dashboard platform).

        Below the description are buttons to navigate to the previous or next topic according to the navigation order. Under that, is a header titled "Community Experiences of Diversity" followed by this text: The stories and themes below are based on community feedback gathered during Phase 2 of the Fostering Diverse Schools project. You can explore the raw feedback (with hyperlink) or visit the project page (with hyperlink) to learn more. Then there is a "share what we missed" button that would allow users to share their own experiences related to the topic.

        Following that, there is a tab bar with three tabs: Student Voices, Staff Voices, and Parent Voices. The first tab is currently selected with a short description that there are 5 stories displayed capturing student voices around diversity.
    }
\end{figure}

We developed an interactive web interface to allow users to explore the 124 generated community stories, as shown in Figure~\ref{fig:platform}.
Given the substantial corpus of stories, we implemented a three-tier navigation system. 
Users can navigate through nine topics via a left sidebar panel, select among three stakeholder perspectives (students, staff, parents) through a tab bar, and progress sequentially through topics using navigation buttons. 
Each topic section begins with a summary explaining the topic's scope and prevalence among all the community responses (not just the 2,480 quotes used as story material).

The interface incorporates several accessibility features to broaden community participation~\cite{toukola_digital_2022, sarangapani_virtualculturalcollaboration_2016}: mobile responsiveness, multilingual support through Google Translate Widget, and content written at the average U.S. adult reading level\footnote{According to the Program for the International Assessment of Adult Competencies (PIAAC), the average American adult reads at a 7th- to 8th-grade level \hyperlink{https://www.nu.edu/blog/49-adult-literacy-statistics-and-facts/}{(source)}.}. 
In addition, when entering the interface, users are directed to a tutorial page that explains the project context and interface mechanics through a manually selected example story.

\begin{figure}[t]
    \centering
    \includegraphics[width=\linewidth]{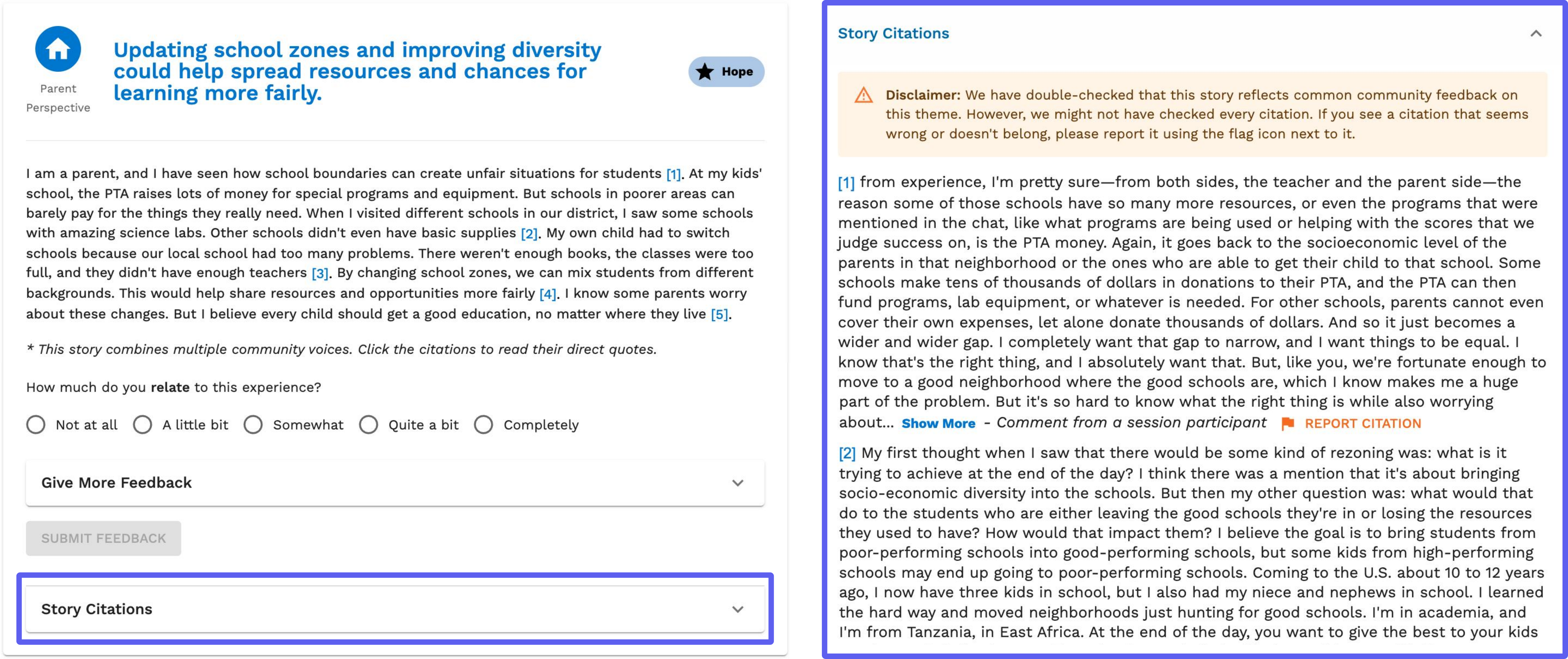}
    \caption{Individual story component including a thematic title, narrative summary, and citations.}
    \label{fig:story}
    \Description{
        Image shows a screenshot of a single narrative story component in the StorySharer interface. This story was pulled from one of three parent voices on diversity and was shown to participants of the user study. The top of the story component first displays, on the very left, a house icon in a blue circle and then the text "Parent Perspective". To the right of the icon, is the thematic story title of "Updating school zones and improving diversity could help spread resources and chances for learning more fairly.". To the very right of the top is a chip with "Hope" written inside. Upon hover, a tooltip would appear, defining hopes as "Hope for future change or improvement".

        Below the story title is a one-paragraph narrative summary with inline citations. This example has the following text: I am a parent, and I have seen how school boundaries can create unfair situations for students [1]. At my kids' school, the PTA raises lots of money for special programs and equipment. But schools in poorer areas can barely pay for the things they really need. When I visited different schools in our district, I saw some schools with amazing science labs. Other schools didn't even have basic supplies [2]. My own child had to switch schools because our local school had too many problems. There weren't enough books, the classes were too full, and they didn't have enough teachers [3]. By changing school zones, we can mix students from different backgrounds. This would help share resources and opportunities more fairly [4]. I know some parents worry about these changes. But I believe every child should get a good education, no matter where they live [5]. The numbers are clickable, and, if clicked on, would navigate to the corresponding citation in the "Story Citations" accordion window. Below the story text is a disclaimer saying "* This story combines multiple community voices. Click the citations to read their direct quotes."

        Below the disclaimer, there is a 5-point Likert-scale question asking "How much do you relate to this experience?" with answer choices "Not at all", "A little bit", "Somewhat", "Quite a bit", and "Completely". Then there is a collapsed accordion titled "Give More Feedback" that when expanded would display three more Likert-scale questions measuring understanding, perceived value, and trust towards the story. Then there is a "Submit Feedback" button that is disabled until at least one Likert-scale question is answered.

        Below the feedback section is the "Story Citations" accordion. When expanded, a disclaimer message is displayed at the top stating: "Disclaimer: We have double-checked that this story reflects common community feedback on this theme. However, we might not have checked every citation. If you see a citation that seems wrong or doesn't belong, please report it using the flag icon next to it.". Then there are two citations listed out. The first part of each citation is displayed followed by a blue "show more" button and then text on where the comment is from and a "report citation" button. 
    }
\end{figure}

As shown in Figure~\ref{fig:story}, individual stories act as self-contained components featuring: (1) stakeholder types, (2) thematic titles, (3) categorical labels (hope, concern, plus), and (4) narrative text with integrated inline citations. 
Each citation links to expandable source material enabling users to verify the story's alignment to real community voices.
Users can report issues with specific citations, along with general feedback on the story through Likert-scale questions measuring relatability, understanding, respect, and trust.


The system was implemented as a containerized web application using React\footnote{\url{https://react.dev/}.} for the front-end interface and Flask\footnote{\url{https://flask.palletsprojects.com/en/2.2.x/}.} for the backend API, deployed on Amazon Web Services\footnote{\url{https://aws.amazon.com/}.}. 
User interactions and feedback data are captured through comprehensive event tracking and stored in a MongoDB\footnote{\url{https://www.mongodb.com/}.} database.

\section{Evaluation}

To evaluate the effectiveness of our human-AI collaborative narrative synthesis approach, we employed a mixed-methods evaluation combining three complementary methodologies: a four-month field deployment of the StorySharer interface during an active school rezoning process, user studies with 21 community members to gather rich qualitative insights on narrative reception and authenticity, and a pre-registered controlled experiment examining how narrative composition variations affect interpersonal measures of connection, understanding, respect, and trust. 
This multi-pronged approach enables us to assess both real-world usage patterns and community reception in high-stakes civic decision-making contexts, while systematically isolating the effects of specific narrative strategies. 
The field deployment provides ecological validity by capturing how community members naturally interact with AI-generated stories during contentious policymaking; the user studies offer deep qualitative understanding of participant reactions to narrative content and the AI-mediated creation process; and the controlled experiment complements these findings by examining causal relationships between narrative structure and interpersonal outcomes. 
Together, these methodologies illuminate both the potential and limitations of human-AI narrative synthesis systems in civic contexts, addressing our research question of how to effectively summarize large-scale qualitative feedback to foster connection and understanding across diverse community perspectives.

\subsection{Field Deployment}

\subsubsection{Deployment Area}

The school district is among the 100 largest in the USA, with just over 50,000 students. The school district is diverse in several dimensions, which translated into a wide diversity of experiences and perspectives in community feedback. The district has a mix of rural, suburban, and urban areas. 67\% of the district's schools are currently designated as Title~I; about 57\% of the student body is measured by the state as economically disadvantaged. About 20\% of the district's schools have a proportion of economically disadvantaged students within 10\% of this district average. Racially, the district student body is predominantly White, Black, and Hispanic, in roughly equal proportions (32\%, 29\%, and 31\%, respectively). About 20\% of schools are majority White, and 29\% (49\%) of these schools have a proportion of White students more than 10\% above (below) the district average. Politically, the district closely mirrors the U.S. as a whole: it has a slightly left-leaning elected school board, and was nearly equally split across votes for Democratic and Republican candidates during the 2024 U.S. Presidential Election.

In late 2023, the district received a federal planning grant to analyze their existing school attendance boundaries, or the catchment areas that describe which neighborhoods are assigned to which schools. These boundaries have not been substantially changed in over 30 years. The purpose of the project was to identify opportunities to ``redistrict'', or change boundaries, in ways that might reduce socioeconomic segregation between schools, improve transportation efficiency, and make other improvements that could enhance equitable access to quality educational opportunities across the district. 

\subsubsection{Deployment Details}

The StorySharer interface launched in February 2025 as part of the school district's ongoing rezoning process. 
The district formally introduced the platform through a press release, establishing its role as an official community engagement report synthesizing feedback collected during an initial community engagement process.
To maximize community reach, we deployed StorySharer as a component within the district's existing boundary map platform, which displays proposed redistricting scenarios for community review.
This deployment decision eliminated the need to maintain separate promotional campaigns, as the extensive outreach efforts for boundary feedback simultaneously drove traffic to the StorySharer. 
The district employed multiple outreach strategies including direct email and SMS to parents and staff, in-person and virtual facilitated feedback sessions, banners and a project page on the district website, physical fliers, and social media campaigns.
The platform remains active as of this writing and will continue operating at least until the rezoning process concludes in Fall 2025, providing a six-month window for community engagement and longitudinal usage analysis.

\subsubsection{Data Collection and Analysis}

We tracked platform usage through anonymous session IDs maintained through Flask-Session and MongoDB. 
Since session IDs are replaced every time the page refreshes, they do not necessarily represent unique users---an important limitation of our tracking approach. 
The platform collected two types of interaction data: feature usage events (capturing specific user actions) and heartbeat data at 3-second intervals (recording page location, device type, and language settings). 
We also implemented four 5-point Likert scale questions to gather feedback on each story measuring relatability, understanding, respect, and trust. 
Additionally, we included a topic-level ``share what we missed'' feature allowing users to contribute thoughts on missing content and indicate their role in the school district, though this received only one response.
For analysis, we quantitatively examined event tracking and heartbeat data from February to May 2025 to calculate metrics including feature usage frequency, navigation patterns, and session duration.

Though we did not track stakeholder type (e.g., parent, student, staff) for users interacting with the StorySharer interface, it shared a website with a rezoning scenario feedback platform---and we did collect stakeholder data from those submissions. The submissions from February to April, the period of active feedback collection that overlapped with the data collection for our deployment, were 79\% ``Parent'', and another 5\% ``Parent and staff''. It therefore seems likely that the large majority of community report visitors were also parents, though we cannot be sure. 

\subsubsection{Findings}

\begin{figure}[t]
    \centering
    \includegraphics[width=\linewidth]{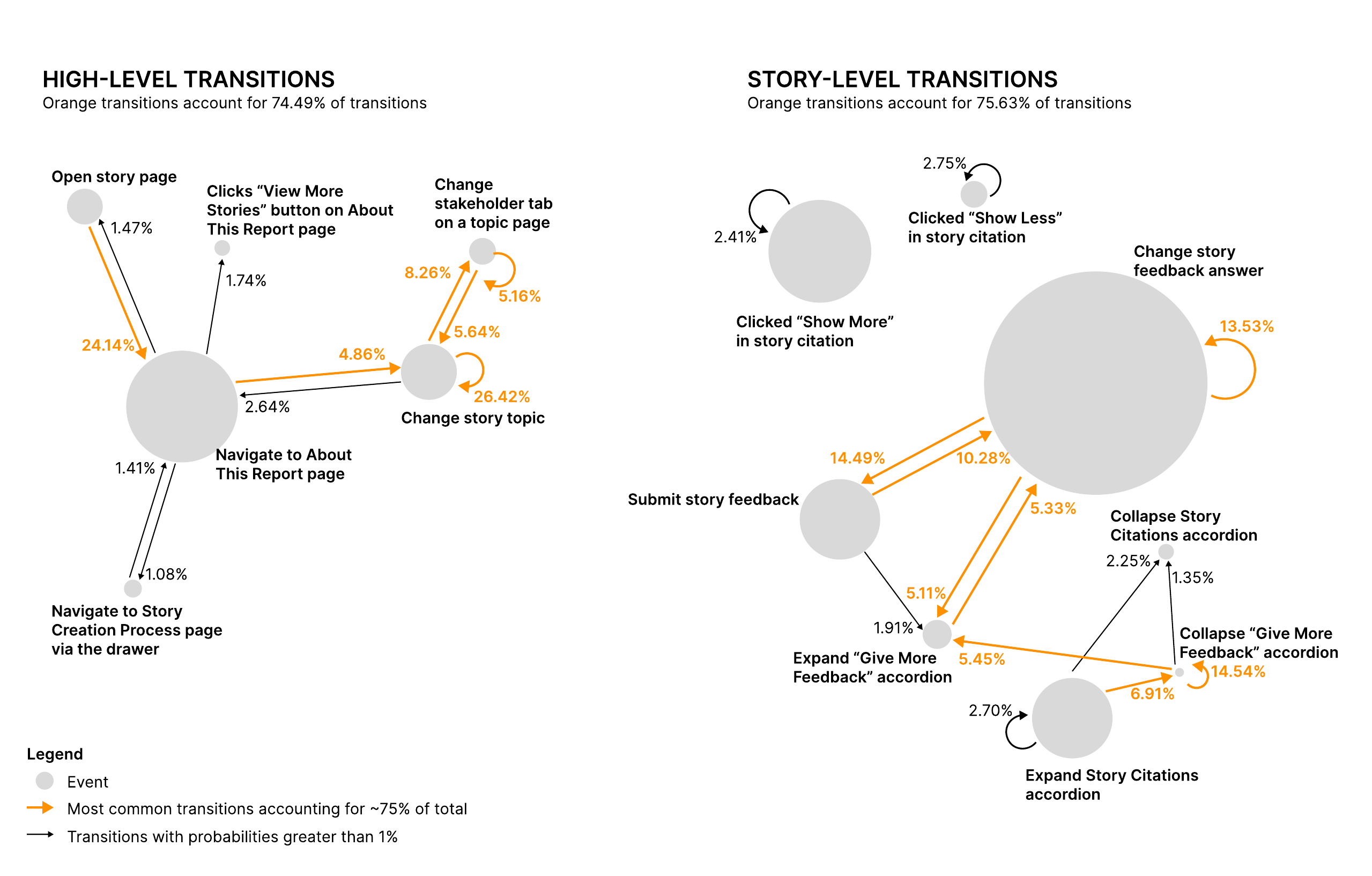}
    \caption{Transitions at the platform level (i.e. high-level transitions) and story level. A node represents a tracked platform event, and has a diameter proportional to the number of times it occurred within its respective ``type'' of transition (high-level or story-level). Orange arrows show the most common transitions, accounting for ~75\% of the total transitions of their respective type.}
    \label{fig:trans}
    \Description{The figure is two node-link diagrams illustrating how participants navigate the platform overall (labelled High-Level Transitions), as well as at the individual story component level (labelled Story-Level Transitions) via tracked events. Each event is represented by a node with a diameter proportional to the number of times it occurred within its respective "type" of transition (high-level or story-level). Arrows between nodes indicate navigation paths, with transition probabilities labeled next to each arrowhead. Only transitions with probabilities greater than 1\% are shown, and the most common transitions accounting for approximately 75\% of the total are highlighted in orange. The most common transitions at the high-level are as follows: Change story topic to Change story topic (26.42\%), Open story page to Navigate to About this Report page (24.14\%), Change story topic to Change stakeholder tab on topic page (8.26\%), Change stakeholder tab on topic page to  Change story topic (5.64\%), Change stakeholder tab on topic page to Change stakeholder tab on topic page (5.16\%), and Navigate to About this Report page to Change story topic (4.86\%). The most common transitions at the story-level are as follows: Collapse “Give More Feedback” accordion to Collapse “Give More Feedback” accordion (14.54\%), Change story feedback answer to Submit story feedback (14.49\%), Change story feedback answer to Change story feedback answer (13.53\%), Submit story feedback to Change story feedback answer (10.28\%), Expand Story Citations accordion to Collapse “Give More Feedback” accordion (6.91\%), Collapse “Give More Feedback” accordion to Expand “Give More Feedback” accordion (5.45\%), Expand “Give More Feedback” accordion to Change story feedback answer (5.33\%), Change story feedback answer to Expand “Give More Feedback” accordion (5.11\%).}
\end{figure}

Our field deployment generated substantial community engagement, with 2,183 sessions visiting the StorySharer interface during the four-month deployment period. 
Of these, 1,675 sessions (77\%) remained active for at least 3 seconds.
Usage patterns revealed high mobile adoption, with 51\% of sessions occurring on mobile devices compared to desktops or tablets. 
Language preferences remained predominantly English, with only 16 sessions (1\%) selecting the Spanish translation option despite the district's multilingual community.
Users demonstrated sustained engagement, spending an average of 6.23 minutes on the platform with high variance (SD = 59.98 minutes), suggesting diverse interaction styles from brief browsing to deep exploration. 
Additionally, participants, who viewed stories, saw an average of 3.13 +/- 6.97 stories per session.
The various interaction styles are highlighted below in Figure~\ref{fig:trans}, which highlights how users flowed through platform-level and story-level actions.
Looking at the story-level transitions, a majority of transitions involve providing feedback on the stories, resulting in 233 responses across 56 sessions (3.3\% of active users).
This concentrated engagement pattern raises questions about what motivated these users to provide feedback: did they believe their input would influence the community engagement process or trusted it would reach decision-makers?
The following user study section contains additional insights from event tracking data, to triangulate community perspectives with observed behaviors.

\subsection{User Study}

\subsubsection{Participants}

\begin{table*}[t]
  \caption{Information about each participant in the user studies (N=21). ``Activity'' refers to self-reported engagement with school district. ``Choice Out'' indicates whether participants currently opted out of their zoned schools. 
  }
  \Description{
    This table presents demographic information for 21 user study participants, organized in 7 columns: ID, Role, Gender, Race/Ethnicity, Income, Activity level, and Choice Out status. Participant composition:
    Roles: 13 parents, 4 staff members, 2 parent/staff members, and 2 "other" community members
    Gender: 18 women (F) and 3 men (M)
    Race/Ethnicity: 12 White participants, 8 Black/African American participants, and 1 who preferred not to disclose
    Income ranges: From \$25,000-49,000 to \$200,000+, with most participants (11) reporting household incomes between \$75,000-149,999. Four participants preferred not to disclose income information.
    Activity levels: Range from "Not active" to "Very active" in school district engagement, with most participants reporting "Active" (N=6) or "Very active" (N=10) involvement
    School choice: 7 participants opted out of their zoned schools ("Yes"), 14 remained in zoned schools ("No")
  }
  \label{tab:participants}
  \small
  \begin{tabular}{@{}lllllll@{}}
    \toprule
    \textbf{ID} & \textbf{Role} & \textbf{Gender} & \textbf{Race/Ethnicity} & \textbf{Income} & \textbf{Activity} & \textbf{Choice Out}\\
    \midrule
    P1 & Parent/Staff & F & Black/African American & \$25-49k & Very active & Yes\\
    P2 & Staff & F & White & \$75-99k & Very active & No\\
    P3 & Parent & M & White & \$150-199k & Very active & No\\
    P4 & Parent & F & White & \$200k+ & Very active & No\\
    P5 & Parent & F & White & \$100-149k & Very active & No\\
    P6 & Other & F & Black/African American & \$200k+ & Not active & No\\
    P7 & Parent & F & Black/African American & Prefer not to say & Active & No\\
    P8 & Staff & F & Black/African American & \$150-199k & Active & No\\
    P9 & Parent & F & Prefer not to say & Prefer not to say & Very active & Yes\\
    P10 & Parent/Staff & F & Black/African American & \$75-99k & Very active & Yes\\
    P11 & Staff & F & Black/African American & \$150-199k & Somewhat active & No\\
    P12 & Parent & F & Black/African American & \$75-99k & Active & No\\
    P13 & Staff & F & White & \$100-149k & Active & No\\
    P14 & Parent & M & White & Prefer not to say & Very active & No\\
    P15 & Parent & F & White & \$100-149k & Active & Yes\\
    P16 & Parent & F & White & \$75-99k & Not active & Yes\\
    P17 & Parent & M & White & Prefer not to say & Prefer not to say & No\\
    P18 & Parent & F & White & \$100-149k & Very active & Yes\\
    P19 & Parent & F & White & \$75-99k & Active & Yes\\
    P20 & Other & F & Black/African American & \$75-99k & Prefer not to say & No\\
    P21 & Parent & F & White & \$100-149k & Very active & No\\
    \bottomrule
  \end{tabular}
\end{table*}

We recruited 21 community members (P1-P21) representing diverse perspectives within the school community: 13 parents, 4 staff members, 2 parents/staff members, and 2 ``other'' community members. 
The sample included 18 women and 3 men, with 12 identifying as White, 8 as Black/African American, and 1 preferring not to disclose race/ethnicity. 
Household incomes ranged from \$25,000 to over \$200,000, with most (n=11) reporting \$75,000-\$149,999.
Participants demonstrated varying district engagement levels from ``not active'' to ``very active'' (i.e., participated in more than five activities/events over the past year). 
Table \ref{tab:participants} summarizes participant attributes.
All participants received \$30 Amazon giftcards as compensation for their time.

\subsubsection{Procedure}

User study sessions lasted 30-60 minutes and were conducted remotely via Zoom.
We began with background questions about participants' school affiliations and familiarity with the district's rezoning efforts. 
Participants then accessed a simplified interface containing tutorial content and two manually selected stories representing contrasting perspectives: one depicting positive parent reactions to rezoning and another covering negative student impacts. 
Stories were presented in randomized order.

Participants read the tutorial and first story naturally without think-aloud protocols while we observed their feature usage and interaction patterns. 
After completing each story, we conducted semi-structured interviews focusing on participant reactions to the story, such as personal relevance (``How does this story relate to your own experiences and perspectives?'') and perceived authenticity (``How authentic or not authentic did the story feel to you?''). 
We repeated this process for the second story, then elicited feedback on specific interface features including thematic titles and story citations.
To assess reactions to AI-generated content, we first asked participants to speculate about the stories' creation process (``How do you think the team came up with these stories?''). 
We then disclosed our AI-assisted pipeline and probed for changes in perception (``How does that change your feelings towards the stories and citations?'').

Finally, we displayed the complete interface showing all 124 stories organized by topic and stakeholder. 
Participants evaluated the system's effectiveness as a community engagement artifact through questions such as ``Is this an effective final report?'' and ``How do the stories make you feel about the overall community engagement process?''
All sessions were audio-recorded and transcribed verbatim with participant consent for subsequent analysis.
See supplementary materials for the questions we asked.

\subsubsection{Data Collection and Analysis}

We analyzed the interview transcripts using inductive thematic analysis~\cite{braun2006using}, which involves an iterative coding process to identify emergent patterns in the data.
First, we segmented the transcripts into paragraphs as units of analysis.
Three authors conducted open coding on a random sample of 7 interviews (33\% of data) to establish initial code categories. 
The research team collaboratively clustered open codes into preliminary categories, developing an initial codebook.
From there, four authors applied the initial codebook to 3 additional interviews to refine definitions and resolve ambiguities. 
Following codebook revision, the same team independently coded 8 more interviews and resolved disagreements through periodic discussions.
This iterative process resulted in 4 codebook revisions over 3 weeks.
The refined codebook has 18 structural codes and 45 thematic codes (view codes in supplementary materials).
Inter-coder reliability reached 0.69 for 11 interviews, measured using Cohen's Kappa~\cite{cohen1960coefficient} and indicating substantial agreement.
One author coded the remaining 10 interviews using the finalized codebook.
The entire coding process took around 28 
hours of independent coding and 7 
hours of live discussion.
After coding was complete, the research team met to condense the codebook into 7 key themes, presented in Table~\ref{tab:themes}.

\subsubsection{Findings}

\begin{table*}[t]
  \caption{Themes emerging from qualitative analysis of community member interviews (N=21). Each theme includes associated codes and a description.}
  \Description{
    This table presents seven major themes identified from user study interviews, organized in three columns: Theme name, Associated Codes, and Description. The themes are:
    Reactions to AI Use - Associated with codes about participants' views of AI and experience with AI. Describes participant reactions to learning about AI-generated content, including their prior AI experience and opinions on AI appropriateness in civic contexts.
    Story Connection - Associated with codes about stories, relatability, understanding, and stakeholder voice. Describes the degree to which participants felt personal connection to story content and identified with represented stakeholder perspectives.
    Authenticity Perception - Associated with codes about stories, citations, AI views, relatability, comprehensibility, and authenticity. Describes factors influencing how participants assessed story authenticity, including citation quality and narrative believability.
    Story Response - Associated with codes about background context, sharing stories/opinions, connections to personal experience, and attitudes toward rezoning. Describes how participants responded to stories by sharing personal experiences, opinions, or connections to their own knowledge.
    Engagement Process Perception - Associated with codes about community engagement process, trust, feeling heard, transparency, and project attitudes. Describes participant perceptions of the broader community engagement process and how the story report contributed to it.
    Interface Interaction - Associated with codes about stories, citations, titles, report layout, user journey, and helpfulness. Describes how participants navigated and interacted with interface features.
    System Improvements - Associated with codes about report layout, content suggestions, UI problems, and comprehensibility issues. Describes participant suggestions for interface improvements, content modifications, and system limitations.
  }
  \label{tab:themes}
  \small 
  \begin{tabular}{@{}>{\raggedright}p{0.10\linewidth}p{0.45\linewidth}p{0.39\linewidth}@{}}
    \toprule
    \textbf{Theme} & \textbf{Associated Codes} & \textbf{Description}\\
    \midrule
    \textbf{Reactions to AI Use} & 
    project view of AI (+/-), general view of AI (+/-), AI boundaries, experience with AI & 
    Participant reactions to disclosure of AI-generated content, including prior AI experience and perceived appropriateness of AI use in civic contexts\\
    \addlinespace
    
    \textbf{Story Connection} & 
    stories, relatability (+/-), understanding (+/-), stakeholder voice (+/-) & 
    Degree to which participants felt personal connection to story content and identified with represented stakeholder perspectives\\
    \addlinespace
    
    \textbf{Authenticity Perception} & 
    stories, citations, project view of AI (+/-), relatability (+/-), incomprehensibility, filling in blanks, authenticity (+/-) & 
    Factors influencing participant assessment of story authenticity, including citation quality and narrative believability\\
    \addlinespace
    
    \textbf{Story Response} & 
    background context, general other, known other, share story, share opinion, unprompted connection to own experience, applicability (+/-), attitude toward rezoning project (+/-) & 
    How participants responded to stories through sharing personal experiences, opinions, or connections to their own knowledge\\
    \addlinespace
    
    \textbf{Engagement Process Perception} & 
    community engagement process, trust (+/-), engagement \mbox{(+/-)}, report (+), feeling heard (+/-), hearing others, continuous engagement (+/-), transparency (+/-), attitude toward rezoning project (+/-) & 
    Participant perceptions of the broader community engagement process and how the story report contributed to the process\\
    \addlinespace
    
    \textbf{Interface Interaction} & 
    stories, citations, title, report layout, user journey, helpfulness (+/-), citation vs story, accuracy (+/-), report (+) & 
    How participants navigated and interacted with interface features\\
    \addlinespace
    
    \textbf{System Improvements} & 
    report layout, content change suggestion, UI problem or suggestion, exposition, incomprehensibility, report (-) & 
    Participant suggestions for interface improvements, content modifications, and system limitations\\
    
    \bottomrule
  \end{tabular}
\end{table*}

\paragraph{Boundaries on AI Involvement and Human Review Foster Trust}
Before the story-generation process was shared, some participants wondered whether AI was involved. P21 said, \emph{``I was assuming it was AI—[it] went through and took all of whatever the responses were, and then just kind of made a paragraph out of it.''} After the process was briefly explained---noting the large dataset and how it was summarized into stories---acceptance of AI usage was conditional yet steady. Participants pointed to the scale of the dataset and the need for help organizing it; as P3 put: \emph{``To me it makes sense. It feels like a pretty good use of resources on it to sift through like an obscene amount of information.''} Knowing that human review and editing were part of the workflow also added confidence. P18 said, \emph{``That feels fine to me [...] I'm glad that you have, like a human person going through and like checking it to make sure you know it kind of rings true to his overall perspective.''}. P6 appreciated the \emph{``subject-matter expertise from humans''} involved in the review. Many participants also drew a clear boundary for use: AI could tag, group, and flag, while people should make the final calls on analysis and wording. P19 summarized: \emph{``You're flagging your individual stories. It's aggregating. And then you're manually editing and changing. And I think to me that's exactly how AI should be used. It's a tool. It's not a replacement. So I don't have any negative reaction to it at all. To me it's using resources smartly.''} 

At the same time, participants noted that the resulting summaries were not only efficient but also effective in representing voices. As P19 highlighted : \emph{``It doesn’t read like a bad AI […] it reads like a true summary of these specifics […] it reads like a human.''} P6 also echoed this sentiment by sharing that she \emph{``find[s] value in the use of AI in very specific instances, and this is one of them.''} Only 5 participants offered measured critiques of using AI in this project context, like P13: \emph{``that's a newer technology [...] I do not know enough about the way that it is to trust it [...] I just don't trust something this important to be done with AI.''}

\paragraph{Relatability and Stakeholder Voice Shape Connection to Stories}

Participants largely used the narrative summaries as a mirror: they reached for personal experience, nearby examples, or familiar community dynamics to make sense of what they read. Several described the scenarios as recognizably ``real''. 
As P6 noted, \emph{``This is a common story from a lot of parents''}.
Relatability often came through recognition, either in oneself or one's social circle. As P5 explained: \emph{``I have friends who are teachers [...] I’ve heard their stories''}. P10 said that little about the narrative felt unfamiliar: \emph{``Not really [anything] stand out. I agree with it on both ends.''}

At the same time, connection was not universal, and several participants distinguished between \emph{relatability} and \emph{understanding}. As P16 shared: \emph{``It didn’t really occur to me as much, because my kids didn’t have that issue [...] That can be a really big issue for some students, and I can understand that.''} Salience also varied by priorities; even while eventually acknowledging that they could \emph{``understand that perspective''}, P18 said, \emph{``I’m more personally affected by academics than [these other concerns].''} In short, participants could follow the logic and stakes even when they did not feel personally ``in'' the story.

Stakeholder voice also shaped the connection. While some valued seeing different perspectives others discounted them. P13 dismissed the importance of student voices saying, \emph{``I know that’s from a student’s perspective, and they may or may not be able to enunciate. [...] But that also is where it comes into being an adult decision that your academic achievement needs to trump you being able to have your athletic needs met.''} Meanwhile, others said student voices were the most important, and P19 highlighted the importance of having multiple voices,  \emph{``I would want to see everything [...]  I’m glad that all voices were accounted for.''} Event-tracking data reinforces this point. Student stories were viewed most frequently (5,389), largely due to their position as the default tab, while parent stories (though positioned after staff) still received far more attention (1,265 views vs 743 views). Parent stories received higher engagement in the form of ratings, with median scores of 5 for relatability, understanding, and perceived value. Parents were the most involved stakeholder group on the deployment website and throughout the rezoning process, so these patterns likely suggest that participants prioritized stories that reflected their own roles, even as they appreciated exposure to other perspectives.

\paragraph{Relatability and Narrative Structure Influence Perceived Authenticity}
Seventeen participants found the narrative summaries and citations authentic, while eight expressed concerns, typically about summaries rather than citations. 
The first-person narrative format enhanced authenticity for some participants, with P19 explaining it made content \emph{``feel less robotic.''} 
Relatability proved strongly connected to authenticity perceptions. 
P1 stated: \emph{``I believe they are real, because they do happen,''} while P2 observed: \emph{``It feels very authentic, if for nothing more than I've heard it before.''} 
Event tracking data confirms this connection, showing that relatability was the most correlated (via Pearson) with trust responses (r = 0.85, N = 81, p-value < .05), followed by value and trust (r = 0.82, N = 85, p-value < .05).

However, authenticity concerns centered on two issues. 
Some found narratives too generic, with P21 explaining: \emph{``I feel like you could pop it and put [it in] a different school district [...] nothing really jumps out as being our specific situation.''} 
Others questioned the ethics of AI-created composite personas, as P3 noted: \emph{``It's like, wait, we've just created an avatar parent.''} 
In addition, event tracking revealed authenticity varied by topic and stakeholder, with diversity receiving lower trust ratings (median 1.5 vs. 5.0) and student/staff stories receiving lower ratings than parent stories (medians of 3.0 and 1.0 vs 5.0).

\paragraph{Narrative Scenes Prompt Narrative Responses}

Participants often responded to stories by sharing narrative details about their own families or experiences.
When asked for their initial thoughts after reading their first story, 12 participants immediately jumped into their own storytelling. For instance, P17 started by saying: \emph{``My wife teaches in the system. I have 3 children in the system.''} A few asked clarifying questions, discussed the UX, or started summarizing the story to gather their thoughts. Only 3 participants (P13, P4, P7) jumped straight into sharing opinions, e.g., \emph{``I know that that happens. And I certainly think that all children should have access to the same resources in schools''} (P4). All but 5 participants shared an unprompted connection to their own experiences at some point during the interview.

Participant storytelling was typically more personal, about people they knew, and frequently prompted by something they found relatable in a story they read. Like P19 shared about a story: \emph{``this is very accurate. So like, I said, we're at a choice school. [shares story about family]. So I think this did a really good job of summarizing how I feel.''} Opinion sharing, by contrast, was more often about generalized groups and concepts, and often about participants' preexisting support or opposition towards the project. P4 shared, \emph{``But you know our concern is not everybody else's kids. Our concern is our kids [...] you know, you have schools that aren't doing well [...] Give those schools the resources they need rather than [rezoning]''}.

The stories presented to participants were a mix of concrete scenes (concrete events and experiences) and themes (underlying interpretations and values). Every participant found something relatable in the scenes of least one of the stories, but reactions to the themes were radically different. Some took new perspectives as food for thought, like P19: \emph{``So that's not something I would have thought about. [...] it's good to get people thinking.''} Others, especially participants who expressed strong preexisting opinions, acknowledged the story scenes as real but then contradicted the story's reasoning, like P13: \emph{``the things that they're talking about are not going to be fixed by redrawing.''}

\paragraph{Perceptions of Engagement Process Reflect Overall Project Views}

Participants had mixed views on the overall community engagement process that informed the report, with 7 expressing only positive views, 4 expressing only negative ones, and 2 expressing both. The report helped give a new perspective to some participants: \emph{``I'm honestly impressed because I didn't realize what all went into it before the maps. And I think a lot of people weren't either so it does help to know [...] that they did try to gather genuine data from parents, families, and staff when coming up with this project and making these decisions''} (P15). Participants' perception of the broader community engagement process (and trust towards the people leading it) generally aligned with their attitudes towards the rezoning project in general.

However, despite fewer participants sharing negative views, they were much more vocal about them: 85\% of critical quotes came from just three participants (P13, P4, and P14), who were also highly critical of the project. Interestingly, these three also talked the least about the sample stories in their interviews (2-3 times fewer quotes than others). These participants generally felt ignored by the process---\emph{``It's like, Okay, we'll hear what your gripes are. Throw them in a closet. Shut the door.''} (P4)---and may have seen  the interview primarily as a chance to give more feedback about the rezoning project. P14 expressed this motivation explicitly: \emph{``That's why I want to talk to you because you're gonna write down what I say.''} P13 and P4 also expressed frustration in keeping up with the iterative feedback process, in which the district shared ongoing learnings from the community: \emph{``I don't have time to continuously sit down and spend this much time analyzing little changes you may have made.''} (P13).

\paragraph{Engagement Depth and Navigation Patterns in the StorySharer Interface}
Participants appreciated the StorySharer's multi-level design, which enabled different levels of engagement with story content.
Participants found thematic titles helpful to \emph{``mentally prepare''} (P2).
P19 highlighted the value of granularity: \emph{``I love that it has these specific [citations]. But then, also, just this is the generative of all everything together [...] I think this is very clear, very concise, but if you want to get into detail, you can, which I appreciate for depending on how much time you have and how much of a concern it is for you.''}
In fact, citation usage patterns varied significantly. 
Some participants like P12 admitted they \emph{``probably wouldn't look at [citations] unless [they...] needed more information.''} 
This aligns with usage analytics showing only 194 of 1,675 sessions (11.58\%) interacted with citations. 
However, a minority of users valued citations over stories, with P21 saying that they \emph{``felt a lot more authentic [...and] a lot more concrete.''} 

Beyond story components, 15 participants appreciated the organizational structure by topics and stakeholders. 
P9 observed: \emph{``This makes it easy [...] People can go and look at the things that really impact them, and see if other people are having the same issue [...] you can see the complaints of the teachers versus the students and the parents.''} 
Figure~\ref{fig:trans} confirms this pattern, with topic and stakeholder navigation representing around 50\% of platform-level transitions. 
However, engagement depth remained limited, with only 550 sessions (32.84\%) progressing beyond the landing page, averaging just 2.75 +/- 1.99 topics per session. 
P5's assessment captures this tension: \emph{``I really like the way that it's laid out [...] although I'm not sure how much people are gonna be actually looking at it.''} 
These findings suggest that while users valued the organizational structure, content volume created barriers for wide-ranging exploration.

\paragraph{Suggested Improvements on Accessibility and Story Presentation}
Participants identified information overload as a primary concern, with the 124-story corpus proving overwhelming despite effective organization in the StorySharer. 
P10 explained: \emph{``I like qualitative data. But it's too much [...] it would be better for me to say, okay, 75\% of the people said that transportation was an issue.''} 
This finding is supported by usage analytics showing users reached the bottom of the topic pages only 36\% of the time (855/2355).
Three participants suggested supplementing narratives with quantitative summaries.
P6 thought the citations helped alleviate a concern that the \emph{``comments were cherry picked''} but confessed that \emph{``nobody's gonna click on those [citations]''}.
Participants also highlighted confusion with the first-person narrative format when stories combined multiple voices. P3 noted: \emph{``The first person probably throws me off a little bit since it is not a single person [...] I feel like it would be interesting to have bulleted lists of direct quotes.''} 

\subsection{Experiment}

To complement the field study and user interviews, we ran a preregistered online experiment (AsPredicted \#243092) to systematically test how different narrative framings shaped interpersonal responses, stance, and collective consideration. We considered four framing strategies: \textit{Scene-dominant} stories (Experience group) that emphasized concrete lived experiences, \textit{Theme-dominant} stories (Opinion group) that emphasized abstract opinions and judgments, \textit{Mixed} stories (Mixed group) that blended experiential detail with thematic reflection, and \textit{Raw Excerpts} (Control group) that presented minimally edited community responses as a control.

The field deployment used \textit{Mixed} stories, which the team selected to provide a balance of experiences and opinions in the community feedback. We considered the other framings and ran a brief pilot study to ``sanity check'' this choice, but had to make a decision before full evaluation to respect deadlines in the ongoing community engagement process. This field deployment and later user interviews surfaced promising signs that these stories could help participants relate to one another despite differences in experience and opinions. However, to better understand how future civic processes could use narrative summaries like these, we wanted to understand how other kinds of story framings might impact community engagement.
The experiment addressed these questions through three preregistered hypotheses:

\textbf{H1: Interpersonal Measures.} 
\begin{itemize}
    \item H1a: Scene-dominant narratives will lead to stronger interpersonal outcomes (connection, understanding, respect, trust, curiosity) compared to theme-dominant narratives.  
    \item H1b: Mixed narratives will show higher effects on interpersonal outcomes than both scene-dominant and theme-dominant narratives.  
\end{itemize}

\textbf{H2: Focus of Consideration.} 
\begin{itemize}
    \item H2a: Scene-dominant narratives will produce larger increases in community-dominant consideration (pre to post) than theme-dominant narratives.  
    \item H2b: Mixed narratives will show higher increases in community-dominant consideration than both scene-dominant and theme-dominant narratives.  
\end{itemize}

\textbf{H3: Stance Change.} 
\begin{itemize}
    \item H3a: Scene-dominant narratives will produce larger changes in stance compared to theme-dominant narratives.  
    \item H3b: Mixed narratives will produce intermediate changes in stance between scene-dominant and theme-dominant narratives.  
\end{itemize}

By testing these hypotheses, we sought to extend the broader goals of our work: to identify narrative strategies that not only support mutual understanding and trust in the moment, but also encourage participants to think beyond their own interests toward the collective, and to examine whether narratives can shift positions in contentious civic debates.

\subsubsection{Participants}
We recruited 200 parents/guardians of U.S. public school students through the Prolific platform. All participants were located in the United States and met the eligibility criteria of having their youngest child born between 2006 and 2019, meaning they were currently enrolled in K–12 education. A minimum response success rate of 98\% was required to ensure data quality. They were compensated at a rate of \$15 per hour, which amounted to \$3.75 for completing the 15-minute study. Our target sample size was powered to detect medium effects, with oversampling to account for preregistered exclusion rules such as insufficient time on task or uniform responses.

\subsubsection{Procedure}
Each participant read four rezoning-related stories drawn from our dataset and rewritten according to their assigned condition. To ensure consistency and relevance, we selected only \textbf{parent stories}, since parents were the most engaged stakeholder group during the rezoning process. Stories were presented in two pairs, each containing one negative and one positive story. Pair~1 addressed how increased diversity was perceived to influence \textbf{education quality} and \textbf{access to opportunity}, while Pair~2 focused on its impact on students’ \textbf{sense of belonging} and \textbf{safety}. The negative stories reflected community voices opposed to updating school zone boundaries, while the positive stories conveyed supportive voices. These topics directly aligned with our preregistered stance measures, enabling us to test whether framing effects held across opposing perspectives on the same issues. Before reading, participants reported their baseline \textbf{stance}, indicating whether they supported or opposed each of the four rezoning topics, and their \textbf{focus of consideration}, indicating whether they viewed the issue more in personal terms or from the perspective of the broader community. Importantly, all stance questions referred to rezoning in participants’ own school districts, not a hypothetical scenario. After each story, they rated five preregistered interpersonal outcomes: \textbf{understanding} (how much they felt they understood the perspectives shaping the story), \textbf{personal connection} (how connected they felt to the voices represented), \textbf{respect} (how much they valued the perspective shared), \textbf{trust} (how much they trusted the voices represented), and \textbf{curiosity} (how much they would like to speak with the people whose voices were included). After each pair of positive and negative stories, participants again reported their stance on the related topics. At the end of the study, participants reported whether they considered the issue primarily in personal terms or from the perspective of the broader community. They then rated the quality of the stories, indicated whether they believed the stories were written by AI, and provided an open-ended reflection on how this affected their perceptions. Full item wording is provided in \Cref{app:experiments}.  

\subsubsection{Data Collection and Analysis}
We analyzed participant-level averages across the four narratives to test condition effects on each outcome. Our preregistered plan specified one-way ANOVAs with Tukey-adjusted pairwise comparisons, complemented by mediation analyses to examine whether interpersonal outcomes accounted for stance and consideration shifts. To check robustness, we also fit mixed-effects models at the story level with random intercepts for participant and story. 
Following preregistered exclusion rules, two participants in the \textit{Theme-dominant} group were removed for failing attention and engagement criteria, yielding 198 participants in the final dataset. Analyses were conducted in Python, and significance was assessed at $\alpha = .05$ with corrections for multiple comparisons where applicable.

\subsubsection{Findings}

Narrative composition had a statistically significant impact on some interpersonal outcomes. Participants who read \textit{scene-dominant} (experience-heavy) stories reported higher \textbf{respect} and \textbf{trust} toward voices with different perspectives than those who read \textit{theme-dominant} (opinion-heavy) stories. For respect, the overall ANOVA was significant, $F(3,194)=5.02, p=.002, \eta^2=.072$, with Tukey tests showing that both \textit{scene-dominant} ($M=3.72$) and \textit{mixed} ($M=3.77$) stories elicited greater respect than \textit{theme-dominant} stories ($M=3.29$; $p=.011$ and $p=.007$, respectively). For trust, the effect was also significant, $F(3,194)=3.02, p=.031, \eta^2=.045$, with scene-dominant stories rated higher ($M=3.64$) than theme-dominant stories ($M=3.26$, $p=.038$). These effect sizes ($\eta^2 = .072$ for respect, $\eta^2 = .045$ for trust) indicate that narrative framing explained 5--7\% of the variance, corresponding to a 0.4--0.5 point difference on the 5-point scale, a small to moderate shift meaningful in brief civic engagement settings. These findings partially support our preregistered hypotheses: consistent with H1a, experience-grounded narratives fostered stronger respect and trust than opinion-dominant ones, while H1b received limited support, with mixed stories showing an advantage only for respect. For downstream outcomes, neither \textbf{stance change} nor \textbf{shifts in focus of consideration} varied significantly by condition (overall ANOVAs were non-significant, all $p > .12$), contrary to preregistered expectations (H2a/b, H3a/b). See \Cref{app:experiment supports} for additional supporting analyses.

Overall, the experiment shows that narrative strategies emphasizing lived experiences are more effective at cultivating respect and trust than those dominated by opinions (partially supporting H1a and H1b), even though they did not directly shift stances or focus of consideration at the condition level (not supporting H2 or H3). These findings align with our user study, where participants described experiential accounts as more ``authentic'' and easier to follow, often responding by sharing their own stories in turn. Both studies suggest that narrative concreteness supports interpersonal respect and trust, even when perspectives differ.

Participants were also asked if they thought the stories were AI-generated, then told that they were (or in the case of \textit{Control}, weren't) and asked if the knowledge changed their feelings. A substantial minority initially suspected that responses were AI generated, with \textit{Control} the lowest (28.0\%), then \textit{Opinion} (32.7\%), \textit{Experience} (38.8\%) and \textit{Mixed} the highest (48.0\%). This high rate of AI suspicion for \textit{Mixed} stories may be part of why they failed to inspire the trust we hypothesized in H1b.
When informed that the (non-control) stories were AI generated, reactions varied widely. Many were unconcerned, because the narrative summaries still felt typical or real to them; or acknowledged lower credibility but continued usefulness, because the fake voices still captured realistic views. Several expressed feeling upset, however, especially in the \textit{Experience} or \textit{Mixed} groups, leading them to devalue the stories or even feel scammed. See \Cref{app:experiment ai reactions} for related participant quotes.

These reactions to AI usage were notably different from the user studies. Several experiment participants expressed significant aversion or upset after learning that the stories were AI-written, while user study participants were broadly accepting---only a few gave measured critiques. There are several possible factors at play here: experiment participants may have been more expressive because they were typing anonymous responses rather than talking with an involved researcher; or experiment participants may have thought that the citations were also generated, rather than just the narrative summaries. However, one important likely factor is the difference between the real civic context versus the experimental sandbox~\cite{horvath2023citizens,eriksson2025makes}. In user studies, participants were also community members, and were made aware of the context, process (e.g., human oversight), and large scale of the feedback in their community. Their acceptance of AI often referred to these factors as key justifications. Plus, many of these community members were interested and engaged with what their community had to say, and often expressed appreciation that the narrative summaries helped make that possible. Experiment participants lacked this context and connection.

\section{Discussion}
Our work explores the research question of how to effectively summarize large-scale qualitative feedback data to foster connection and understanding across diverse community perspectives.
Through the StoryBuilder pipeline and StorySharer interface, we transformed 2,480 community experiences into 124 accessible narratives and deployed them during an active, contentious school rezoning process---a high-stakes policymaking domain notorious for polarization and community conflict~\cite{gillani2023air}. 
This real-world deployment context distinguishes our work from controlled studies, revealing how AI-mediated narrative synthesis performs under the pressures of actual civic decision-making where trust is fragile and stakes are high. 
Our mixed-methods evaluation demonstrates that narrative summaries helped community members relate across diverse perspectives, even when stories conflicted with readers' experiences. 
However, emotional connection proved more complex, varying primarily with story relatability: in both the field deployment analytics and user studies, parents seemed more likely to feel connected to the stories shared by other parents, and less so to those shared by students and staff in the district.
An experiment controlling for relatability (by recruiting only parents and showing parent perspectives) revealed that experience-heavy narratives fostered greater trust and respect than opinion-focused stories.
Some experiment participants were strongly skeptical about AI usage for writing the narrative summaries, but user study participants in the real civic context broadly found it justified and helpful.
This work contributes three key advances to HCI and civic technology: a scalable human-AI collaborative pipeline for narrative synthesis, empirical insights from a field deployment during a contentious policymaking process, and controlled experimental evidence on how narrative composition affects interpersonal measures. We reflect on each of these in greater detail below.

\subsection{Human-AI Collaboration in Civic Narrative Synthesis}

When developing scalable narrative synthesis methods, AI proved critical given the large data volume of 2,480 community experiences ($\sim$358k words) and tight timelines essential for closing community engagement feedback loops. 
Most user study participants acknowledged AI's utility for analyzing large datasets, even those with generally negative AI attitudes. 
However, creating narratives that reflect both diverse and shared community experiences required substantial human oversight. 
Human involvement was especially important during topic-stakeholder categorization and theme generation, as these decisions fundamentally shaped the perspectives presented in final narratives.
Our approach incorporated extensive AI scaffolding through task-specific prompts (e.g., one for theme classification, another for scene extraction from community experiences) to prevent hallucinations observed in initial iterations.
This aligns with prior work demonstrating that specialized AI tools outperform general-purpose systems for specific tasks~\cite{overney2025coalesce}.
Additionally, when reviewing AI-generated stories, we observed challenges in producing and evaluating accurate citations from experiential data. 
This citation accuracy problem extends beyond civic contexts to research literature~\cite{walters2023fabrication, mugaanyi2024evaluation}, highlighting a broader limitation of AI-generated content. 

Regarding AI involvement, the context and reasons for AI use seemed to be important factors in participant reactions \cite{horvath2023citizens,eriksson2025makes}. Experiment participants lacked context and were less accepting of AI use, especially when they read experience-rich stories authored by AI. By contrast, user study participants embedded in the real civic context were broadly accepting, and regularly cited the large scale of the data, value of the output (getting to see what their community said), and involvement of human oversight as reasons to accept the AI usage.
User study participants also raised important concerns about the boundaries between human and AI roles, such as having AI help with analysis but not with story writing.
This echoes established frameworks for human-AI collaboration in related processes, such as data storytelling~\cite{liWhereAreWe2024}, where different roles apply at various sensemaking stages. 
Our StoryBuilder pipeline employed an ``AI creator, human optimizer'' model~\cite{liWhereAreWe2024}, which risks over-reliance when human reviewers lack complementary expertise~\cite{zhang2020effect, burtonAlgorithmAversionHumanMachine2023, gajosPeopleEngageCognitively2022}. 
Our human reviewers included people who possessed deep expertise in the community feedback data and were actively involved in rezoning decisions, enabling effective AI-generated content evaluation.
However, this level of domain expertise may not exist in other contexts.

A particularly complex boundary emerged around AI's role in creating composite personas from multiple community voices.
While some appreciated the personal, accessible tone, others questioned reading multiple experiences as an ``avatar parent'' (P3).
This tension parallels challenges in AI-generated persona development~\cite{GarciaMarza2024Algorithmic, fulay2025empty} and highlights the need for careful testing of AI-mediated representation in civic contexts to establish appropriate boundaries and avoid ethical risks.
The distinction between ``revising something [...] compared to generating something from scratch'' (P13) becomes particularly salient when AI creates synthetic personas that may not represent any individual's actual experience.
Future work could investigate methods for human validation of AI-generated composite narratives, including participatory review processes that involve original feedback contributors in assessing whether synthesized stories accurately represent their experiences.

\subsection{Impact of Narrative Summaries in Real-World Community Engagement}

While the StorySharer's organization of 124 narratives by topic and stakeholder was appreciated by participants, the volume of content created information overload challenges. 
Some participants suggested creating additional topic overview information.
This mirrors the author-driven versus user-driven dichotomy in visualization research~\cite{segel2010narrative, stolper2018data}, where explanatory interfaces prioritize clarity while exploratory tools facilitate open-ended sensemaking.
We attempted to balance between explanatory and exploratory use-cases.
For example, our design of individual story components incorporated multiple levels of granularity: thematic titles, paragraph summaries, and detailed citations, which allowed users to choose their engagement depth based on available time and interest. 
However, usage analytics revealed a critical tension, in which citations were valued by interview participants, yet rarely accessed in the field deployment (only 12\% of users). This echoes broader patterns where source attribution is perceived as essential for AI-generated content but often underutilized in practice, potentially causing overreliance on AI~\cite{schroeder2025forage, si2023large}. 
This raises questions about how to design more accessible ways to encourage citation interaction.
For example, creating an audio medley of citation content that is easier to engage with than clicking through individual inline citations or reading lengthy text blocks \cite{chandler2015listening}.
In addition, several participants suggested integrating quantitative summaries alongside narratives, though prior research indicates narratives may be more effective for reducing exclusionary attitudes and increasing respect~\cite{kubinPersonalExperiencesBridge2021, kallaWhichNarrativeStrategies2023, audette2020personal}, raising questions about how to balance quantitative overview data with qualitative narrative depth.
This design challenge suggests opportunities for future work to explore integrated presentation methods that preserve narrative engagement while meeting users' needs for quick overview information, building on ideas like narrative visualizations~\cite{segel2010narrative, figueiras2014narrative} or process narratives~\cite{freeman2024can, shaffer2013effects}.

Narrative summaries could also help to promote deeper ongoing community engagement, particularly critical in contentious policymaking contexts like school rezoning where traditional approaches often exacerbate polarization and distrust~\cite{gillani2023air}. 
In this project, the stories were used as a final report to summarize the overall results of the community engagement process. 
But in user studies, reading stories consistently inspired community members to reflect on their own personal experiences and share their own stories, even when discussing divisive topics like diversity and school boundaries. 
This suggests that presenting diverse community experiences in story format might be a powerful scaffold for continuing constructive conversations in high-stakes decision-making processes, enabling an iterative engagement model: the community shares feedback and experiences, which are then synthesized into stories, which inspire further story-sharing as community members engage with the summaries. 
Such approaches may prove especially valuable in representative political contexts where civic leaders must navigate deeply held disagreements while maintaining democratic legitimacy and community trust throughout extended decision-making cycles~\cite{overney2025boundarease, castro2022narratives}.

\subsection{Narrative Composition Strategies and Their Impact on Interpersonal Connection}

The controlled experiment confirms that not all narrative forms are equally effective at fostering interpersonal outcomes. Stories emphasizing lived experiences consistently generated higher respect and trust than those emphasizing opinions, with balanced narratives also outperforming opinion-heavy ones on respect. 
These results build on McAdams’ conceptual framework of narratives progressing from concrete scenes to abstract themes \cite{mcadams2001psychology}, and on work showing that sharing personal experiences can be more effective than exchanging opinions in contexts of disagreement, where rational argument often fails to persuade \cite{talbi2024epistemic}. They also align with recent studies demonstrating that experience-grounded accounts can foster understanding and connection across social divides better than facts \cite{kubinPersonalExperiencesBridge2021,hagmann2024personal}.

The user study helps explain why these dynamics occur. Participants frequently described experiential accounts as ``authentic'' and ``accurately depicting what was going on,'' often drawing on their own memories and community knowledge to make sense of the stories. By contrast, when narratives felt generic or overly thematic, participants were more likely to question their authenticity or relevance to the local rezoning context. Citations played an important role in mediating this boundary: they anchored stories to recognizable voices, helping participants trust that accounts reflected real contributions. 
Taken together, these results suggest that narrative strategies emphasizing scenes provide a stronger basis for valuing and trusting perspectives across divides, but that authenticity signals (e.g., citations, clear voice attribution) remain essential in practice.

At the same time, our findings underscore that interpersonal connection does not necessarily translate into perspective change. In both the user study and field deployment, some participants with strongly held views used stories to reinforce rather than revise their positions. They acknowledged the experiential validity of individual scenes but rejected the broader themes, actively ``digging in their heels.'' These patterns suggest a risk that exposure to diverse experiences may sometimes accentuate polarization in contested policy domains, echoing prior work on how exposure to opposing viewpoints can backfire under certain conditions \cite{bail2018polarization}. In our experiment, this dynamic may help explain why respect and trust increased, but downstream measures of stance and collective consideration did not shift. 
These convergences and tensions highlight open questions for future research, which we elaborate in the sections that follow.

\subsection{Risks and Ethical Considerations}
Our deployment of AI-generated narrative summaries in a real-world civic decision-making process revealed several risks and ethical considerations. 
While our StoryBuilder approach incorporated human oversight to mitigate AI-generated inaccuracies and systematic biases, these safeguards may prove insufficient in contexts with limited review capacity, particularly given the resource and time constraints typical of community engagement processes~\cite{irvinCitizenParticipationDecision2004, brysonDesigningPublicParticipation2013}. 
In contentious domains like policymaking to increase school diversity, AI's potential to amplify existing biases toward specific sub-populations~\cite{slattery2024systematic, whose_opinions} could exacerbate rather than bridge societal divides, directly undermining the goal of fostering shared understanding. 
Additionally, our approach raises ethical questions about using non-human entities like AI to analyze and synthesize authentic community voices.
Despite transparency measures including story citations and detailed methodology explanations, important tensions remain about whether AI-mediated narrative synthesis preserves the integrity of human expression or introduces subtle distortions that community members cannot easily detect. 
Future implementations should consider mandatory human validation at critical junctures, particularly those involving vulnerable communities or contentious policy decisions.

\section{Limitations and Future Work}

Several important limitations shape how our findings should be interpreted. 
First, our evaluation was constrained by a tight deployment timeline, preventing systematic review of all 124 generated stories and citations before their release. 
While we manually reviewed every deployed story and included disclaimers about potential citation inaccuracies with user reporting mechanisms, more comprehensive validation would strengthen future implementations. 
Second, our user study participants were recruited from community members who had opted into additional feedback requests, potentially biasing toward more actively engaged individuals. 
We were also unable to interview students, an important stakeholder group in school district contexts. 
Third, our controlled experiment was conducted through Prolific with parents outside the field deployment community, limiting ecological validity compared to testing with community members actively experiencing the rezoning process.

Looking forward, we see several promising directions for future work in narrative construction and validation. 
First, designers of civic narrative systems should carefully balance scenes and themes, leveraging lived experiences to build respect and trust while avoiding overly abstract framings that reduce authenticity. 
Second, systems should attend to authenticity cues through comprehensive citations, transparency about AI authorship, and clarity about whose voices are represented. 
Third, we plan to investigate more robust validation mechanisms for AI-generated civic content, particularly developing automated approaches to verify citation accuracy. 
Finally, future research could explore how narrative strategies interact with participants' prior attitude strength, as our results suggest that interpersonal respect may be necessary but insufficient for collective perspective-taking, raising questions about how narrative design and deliberative contexts can support constructive stance shifts.

We also see value in expanding evaluation methodologies and deployment contexts. 
First, we aim to expand the StoryBuilder and StorySharer approach to other civic contexts beyond school districts, including municipal governments and organizations conducting constituency-listening processes. 
This expansion raises important questions about how narrative synthesis strategies transfer across different community engagement domains and decision-making structures. 
Lastly, we see value in longitudinal studies examining how repeated exposure to AI-mediated narrative synthesis affects trust, participation patterns, and collective understanding in iterative community engagement, while exploring stories' potential as active conversation facilitators rather than passive summaries.

\section{Conclusion}

This study demonstrates how a novel human-AI collaborative narrative synthesis method can augment community engagement in representative political contexts, addressing the critical challenge of maintaining shared understanding between civic leaders and constituents at scale. 
Through the StoryBuilder pipeline and StorySharer interface, we condensed 2,480 community experiences into 124 narrative summaries that enabled community members to acknowledge different, sometimes opposing, viewpoints.
However, our findings also reveal mixed perspectives on the authenticity of AI-generated content, the quality and scale of information represented in the stories, and whose voices are deemed to be most relevant and resonant for making sense of community sentiments.
Overall, our results highlight both the promise and limitations of narrative synthesis in civic contexts. Future advances in civic human-centered AI require careful attention to human–AI role boundaries, systematic validation mechanisms for AI-generated content, and longitudinal evaluation of narrative strategies across different community engagement settings.


\bibliographystyle{ACM-Reference-Format}
\bibliography{main}


\begin{thebibliography}{101}


\ifx \showCODEN    \undefined \def \showCODEN     #1{\unskip}     \fi
\ifx \showISBNx    \undefined \def \showISBNx     #1{\unskip}     \fi
\ifx \showISBNxiii \undefined \def \showISBNxiii  #1{\unskip}     \fi
\ifx \showISSN     \undefined \def \showISSN      #1{\unskip}     \fi
\ifx \showLCCN     \undefined \def \showLCCN      #1{\unskip}     \fi
\ifx \shownote     \undefined \def \shownote      #1{#1}          \fi
\ifx \showarticletitle \undefined \def \showarticletitle #1{#1}   \fi
\ifx \showURL      \undefined \def \showURL       {\relax}        \fi
\providecommand\bibfield[2]{#2}
\providecommand\bibinfo[2]{#2}
\providecommand\natexlab[1]{#1}
\providecommand\showeprint[2][]{arXiv:#2}

\bibitem[Abram et~al\mbox{.}(2020)]%
        {abram2020methods}
\bibfield{author}{\bibinfo{person}{Marissa~D Abram}, \bibinfo{person}{Karen~T Mancini}, {and} \bibinfo{person}{R~David Parker}.} \bibinfo{year}{2020}\natexlab{}.
\newblock \showarticletitle{Methods to integrate natural language processing into qualitative research}.
\newblock \bibinfo{journal}{\emph{International Journal of Qualitative Methods}}  \bibinfo{volume}{19} (\bibinfo{year}{2020}), \bibinfo{pages}{1609406920984608}.
\newblock


\bibitem[Arguedas and Simon(2023)]%
        {arguedas2023automating}
\bibfield{author}{\bibinfo{person}{Amy~Ross Arguedas} {and} \bibinfo{person}{Felix~M Simon}.} \bibinfo{year}{2023}\natexlab{}.
\newblock \showarticletitle{Automating democracy: Generative AI, journalism, and the future of democracy}.
\newblock  (\bibinfo{year}{2023}).
\newblock


\bibitem[Audette et~al\mbox{.}(2020)]%
        {audette2020personal}
\bibfield{author}{\bibinfo{person}{Nicole Audette}, \bibinfo{person}{Jeremy Horowitz}, {and} \bibinfo{person}{Kristin Michelitch}.} \bibinfo{year}{2020}\natexlab{}.
\newblock \showarticletitle{Personal narratives reduce negative attitudes toward refugees and immigrant outgroups: Evidence from Kenya}.
\newblock \bibinfo{journal}{\emph{Vanderbilt University Center for the Study of Democratic Institutions, Working Paper}}  \bibinfo{volume}{1} (\bibinfo{year}{2020}).
\newblock


\bibitem[Bail et~al\mbox{.}(2018)]%
        {bail2018polarization}
\bibfield{author}{\bibinfo{person}{Christopher~A. Bail}, \bibinfo{person}{Lisa~P. Argyle}, \bibinfo{person}{Taylor~W. Brown}, \bibinfo{person}{John~P. Bumpus}, \bibinfo{person}{Haohan Chen}, \bibinfo{person}{M.~B.~Fallin Hunzaker}, \bibinfo{person}{Jaemin Lee}, \bibinfo{person}{Marcus Mann}, \bibinfo{person}{Friedolin Merhout}, {and} \bibinfo{person}{Alexander Volfovsky}.} \bibinfo{year}{2018}\natexlab{}.
\newblock \showarticletitle{Exposure to opposing views on social media can increase political polarization}.
\newblock \bibinfo{journal}{\emph{Proceedings of the National Academy of Sciences}} \bibinfo{volume}{115}, \bibinfo{number}{37} (\bibinfo{year}{2018}), \bibinfo{pages}{9216--9221}.
\newblock
\href{https://doi.org/10.1073/pnas.1804840115}{doi:\nolinkurl{10.1073/pnas.1804840115}}
\showeprint{https://www.pnas.org/doi/pdf/10.1073/pnas.1804840115}


\bibitem[Baumer et~al\mbox{.}(2017)]%
        {baumer2017comparing}
\bibfield{author}{\bibinfo{person}{Eric~PS Baumer}, \bibinfo{person}{David Mimno}, \bibinfo{person}{Shion Guha}, \bibinfo{person}{Emily Quan}, {and} \bibinfo{person}{Geri~K Gay}.} \bibinfo{year}{2017}\natexlab{}.
\newblock \showarticletitle{Comparing grounded theory and topic modeling: Extreme divergence or unlikely convergence?}
\newblock \bibinfo{journal}{\emph{Journal of the Association for Information Science and Technology}} \bibinfo{volume}{68}, \bibinfo{number}{6} (\bibinfo{year}{2017}), \bibinfo{pages}{1397--1410}.
\newblock


\bibitem[Beeferman and Gillani(2023)]%
        {beefermanFeedbackMap2023}
\bibfield{author}{\bibinfo{person}{Doug Beeferman} {and} \bibinfo{person}{Nabeel Gillani}.} \bibinfo{year}{2023}\natexlab{}.
\newblock \showarticletitle{FeedbackMap: A Tool for Making Sense of Open-ended Survey Responses}. In \bibinfo{booktitle}{\emph{CSCW '23 Companion: Companion Publication of the 2023 Conference on Computer Supported Cooperative Work and Social Computing}}. \bibinfo{pages}{395--397}.
\newblock
\href{https://doi.org/10.1145/3584931.3607496}{doi:\nolinkurl{10.1145/3584931.3607496}}


\bibitem[Braun and Clarke(2006)]%
        {braun2006using}
\bibfield{author}{\bibinfo{person}{Virginia Braun} {and} \bibinfo{person}{Victoria Clarke}.} \bibinfo{year}{2006}\natexlab{}.
\newblock \showarticletitle{Using thematic analysis in psychology}.
\newblock \bibinfo{journal}{\emph{Qualitative research in psychology}} \bibinfo{volume}{3}, \bibinfo{number}{2} (\bibinfo{year}{2006}), \bibinfo{pages}{77}.
\newblock


\bibitem[Bryson et~al\mbox{.}(2013)]%
        {brysonDesigningPublicParticipation2013}
\bibfield{author}{\bibinfo{person}{John~M. Bryson}, \bibinfo{person}{Kathryn~S. Quick}, \bibinfo{person}{Carissa~Schively Slotterback}, {and} \bibinfo{person}{Barbara~C. Crosby}.} \bibinfo{year}{2013}\natexlab{}.
\newblock \showarticletitle{Designing {{Public Participation Processes}}}.
\newblock \bibinfo{journal}{\emph{Public Administration Review}} \bibinfo{volume}{73}, \bibinfo{number}{1} (\bibinfo{year}{2013}), \bibinfo{pages}{23--34}.
\newblock
\showISSN{1540-6210}
\href{https://doi.org/10.1111/j.1540-6210.2012.02678.x}{doi:\nolinkurl{10.1111/j.1540-6210.2012.02678.x}}


\bibitem[Burton et~al\mbox{.}(2023)]%
        {burtonAlgorithmAversionHumanMachine2023}
\bibfield{author}{\bibinfo{person}{Jason~W. Burton}, \bibinfo{person}{Mari-Klara Stein}, {and} \bibinfo{person}{Tina~Blegind Jensen}.} \bibinfo{year}{2023}\natexlab{}.
\newblock \showarticletitle{Beyond {{Algorithm Aversion}} in {{Human-Machine Decision-Making}}}.
\newblock In \bibinfo{booktitle}{\emph{Judgment in {{Predictive Analytics}}}}, \bibfield{editor}{\bibinfo{person}{Matthias Seifert}} (Ed.). \bibinfo{publisher}{Springer International Publishing}, \bibinfo{address}{Cham}, \bibinfo{pages}{3--26}.
\newblock
\showISBNx{978-3-031-30085-1}
\href{https://doi.org/10.1007/978-3-031-30085-1_1}{doi:\nolinkurl{10.1007/978-3-031-30085-1_1}}


\bibitem[Carlsen and Ralund(2022)]%
        {carlsen2022computational}
\bibfield{author}{\bibinfo{person}{Hjalmar~Bang Carlsen} {and} \bibinfo{person}{Snorre Ralund}.} \bibinfo{year}{2022}\natexlab{}.
\newblock \showarticletitle{Computational grounded theory revisited: From computer-led to computer-assisted text analysis}.
\newblock \bibinfo{journal}{\emph{Big Data \& Society}} \bibinfo{volume}{9}, \bibinfo{number}{1} (\bibinfo{year}{2022}), \bibinfo{pages}{20539517221080146}.
\newblock


\bibitem[Carroll(1997)]%
        {carroll1997scenario}
\bibfield{author}{\bibinfo{person}{John~M Carroll}.} \bibinfo{year}{1997}\natexlab{}.
\newblock \showarticletitle{Scenario-based design}.
\newblock In \bibinfo{booktitle}{\emph{Handbook of human-computer interaction}}. \bibinfo{publisher}{Elsevier}, \bibinfo{pages}{383--406}.
\newblock


\bibitem[Castro et~al\mbox{.}(2022)]%
        {castro2022narratives}
\bibfield{author}{\bibinfo{person}{Andrene~J Castro}, \bibinfo{person}{Genevieve Siegel-Hawley}, \bibinfo{person}{Kimberly Bridges}, {and} \bibinfo{person}{Shenita~E Williams}.} \bibinfo{year}{2022}\natexlab{}.
\newblock \showarticletitle{Narratives of race in school rezoning: How the politics of whiteness shape belonging, leadership decisions, and school attendance boundaries}.
\newblock \bibinfo{journal}{\emph{AERA Open}}  \bibinfo{volume}{8} (\bibinfo{year}{2022}), \bibinfo{pages}{23328584221091274}.
\newblock


\bibitem[Chandler et~al\mbox{.}(2015)]%
        {chandler2015listening}
\bibfield{author}{\bibinfo{person}{Rasheeta Chandler}, \bibinfo{person}{Erica Anstey}, {and} \bibinfo{person}{Henry Ross}.} \bibinfo{year}{2015}\natexlab{}.
\newblock \showarticletitle{Listening to Voices and Visualizing Data in Qualitative Research: Hypermodal Dissemination Possibilities}.
\newblock \bibinfo{journal}{\emph{SAGE Open}} \bibinfo{volume}{5}, \bibinfo{number}{2} (\bibinfo{year}{2015}), \bibinfo{pages}{2158244015592166}.
\newblock
\href{https://doi.org/10.1177/2158244015592166}{doi:\nolinkurl{10.1177/2158244015592166}}
\showeprint{https://doi.org/10.1177/2158244015592166}


\bibitem[Chandrasekar et~al\mbox{.}(2024)]%
        {chandrasekar2024making}
\bibfield{author}{\bibinfo{person}{Abinaya Chandrasekar}, \bibinfo{person}{Sigr{\'u}n~Eyr{\'u}nard{\'o}ttir Clark}, \bibinfo{person}{Sam Martin}, \bibinfo{person}{Samantha Vanderslott}, \bibinfo{person}{Elaine~C Flores}, \bibinfo{person}{David Aceituno}, \bibinfo{person}{Phoebe Barnett}, \bibinfo{person}{Cecilia Vindrola-Padros}, {and} \bibinfo{person}{Norha Vera San~Juan}.} \bibinfo{year}{2024}\natexlab{}.
\newblock \showarticletitle{Making the most of big qualitative datasets: a living systematic review of analysis methods}.
\newblock \bibinfo{journal}{\emph{Frontiers in big Data}}  \bibinfo{volume}{7} (\bibinfo{year}{2024}), \bibinfo{pages}{1455399}.
\newblock


\bibitem[Charmaz(2006)]%
        {charmaz2006constructing}
\bibfield{author}{\bibinfo{person}{Kathy Charmaz}.} \bibinfo{year}{2006}\natexlab{}.
\newblock \bibinfo{booktitle}{\emph{Constructing grounded theory: A practical guide through qualitative analysis}}.
\newblock \bibinfo{publisher}{sage}.
\newblock


\bibitem[Chen et~al\mbox{.}(2021)]%
        {chen2021graphplan}
\bibfield{author}{\bibinfo{person}{Hong Chen}, \bibinfo{person}{Raphael Shu}, \bibinfo{person}{Hiroya Takamura}, {and} \bibinfo{person}{Hideki Nakayama}.} \bibinfo{year}{2021}\natexlab{}.
\newblock \showarticletitle{GraphPlan: Story generation by planning with event graph}.
\newblock \bibinfo{journal}{\emph{arXiv preprint arXiv:2102.02977}} (\bibinfo{year}{2021}).
\newblock


\bibitem[Cohen(1960)]%
        {cohen1960coefficient}
\bibfield{author}{\bibinfo{person}{Jacob Cohen}.} \bibinfo{year}{1960}\natexlab{}.
\newblock \showarticletitle{A coefficient of agreement for nominal scales}.
\newblock \bibinfo{journal}{\emph{Educational and psychological measurement}} \bibinfo{volume}{20}, \bibinfo{number}{1} (\bibinfo{year}{1960}), \bibinfo{pages}{37--46}.
\newblock


\bibitem[Corbett and Le~Dantec(2018)]%
        {corbettProblemCommunityEngagement2018}
\bibfield{author}{\bibinfo{person}{Eric Corbett} {and} \bibinfo{person}{Christopher~A. Le~Dantec}.} \bibinfo{year}{2018}\natexlab{}.
\newblock \showarticletitle{The {{Problem}} of {{Community Engagement}}: {{Disentangling}} the {{Practices}} of {{Municipal Government}}}. In \bibinfo{booktitle}{\emph{Proceedings of the 2018 {{CHI Conference}} on {{Human Factors}} in {{Computing Systems}}}} \emph{(\bibinfo{series}{{{CHI}} '18})}. \bibinfo{publisher}{Association for Computing Machinery}, \bibinfo{address}{New York, NY, USA}, \bibinfo{pages}{1--13}.
\newblock
\showISBNx{978-1-4503-5620-6}
\href{https://doi.org/10.1145/3173574.3174148}{doi:\nolinkurl{10.1145/3173574.3174148}}


\bibitem[Dahlstrom(2014)]%
        {dahlstrom2014using}
\bibfield{author}{\bibinfo{person}{Michael~F Dahlstrom}.} \bibinfo{year}{2014}\natexlab{}.
\newblock \showarticletitle{Using narratives and storytelling to communicate science with nonexpert audiences}.
\newblock \bibinfo{journal}{\emph{Proceedings of the national academy of sciences}} \bibinfo{volume}{111}, \bibinfo{number}{supplement\_4} (\bibinfo{year}{2014}), \bibinfo{pages}{13614--13620}.
\newblock


\bibitem[Drouhard et~al\mbox{.}(2017)]%
        {drouhard2017aeonium}
\bibfield{author}{\bibinfo{person}{Margaret Drouhard}, \bibinfo{person}{Nan-Chen Chen}, \bibinfo{person}{Jina Suh}, \bibinfo{person}{Rafal Kocielnik}, \bibinfo{person}{Vanessa Pena-Araya}, \bibinfo{person}{Keting Cen}, \bibinfo{person}{Xiangyi Zheng}, {and} \bibinfo{person}{Cecilia~R Aragon}.} \bibinfo{year}{2017}\natexlab{}.
\newblock \showarticletitle{Aeonium: Visual analytics to support collaborative qualitative coding}. In \bibinfo{booktitle}{\emph{2017 IEEE Pacific Visualization Symposium (PacificVis)}}. IEEE, \bibinfo{pages}{220--229}.
\newblock


\bibitem[Eriksson et~al\mbox{.}(2025)]%
        {eriksson2025makes}
\bibfield{author}{\bibinfo{person}{Kimmo Eriksson}, \bibinfo{person}{Simon Karlsson}, \bibinfo{person}{Irina Vartanova}, {and} \bibinfo{person}{Pontus Strimling}.} \bibinfo{year}{2025}\natexlab{}.
\newblock \showarticletitle{What Makes AI Applications Acceptable or Unacceptable? A Predictive Moral Framework}.
\newblock \bibinfo{journal}{\emph{arXiv preprint arXiv:2508.19317}} (\bibinfo{year}{2025}).
\newblock


\bibitem[Figueiras(2014)]%
        {figueiras2014narrative}
\bibfield{author}{\bibinfo{person}{Ana Figueiras}.} \bibinfo{year}{2014}\natexlab{}.
\newblock \showarticletitle{Narrative visualization: A case study of how to incorporate narrative elements in existing visualizations}. In \bibinfo{booktitle}{\emph{2014 18th international conference on information visualisation}}. IEEE, \bibinfo{pages}{46--52}.
\newblock


\bibitem[Freeman et~al\mbox{.}(2024)]%
        {freeman2024can}
\bibfield{author}{\bibinfo{person}{Alexandra~LJ Freeman}, \bibinfo{person}{Lisa-Maria Tanase}, \bibinfo{person}{Claudia~R Schneider}, {and} \bibinfo{person}{John Kerr}.} \bibinfo{year}{2024}\natexlab{}.
\newblock \showarticletitle{Can narrative help people engage with and understand information without being persuasive? An empirical study}.
\newblock \bibinfo{journal}{\emph{Royal Society open science}} \bibinfo{volume}{11}, \bibinfo{number}{7} (\bibinfo{year}{2024}), \bibinfo{pages}{231708}.
\newblock


\bibitem[Fulay and Roy(2025)]%
        {fulay2025empty}
\bibfield{author}{\bibinfo{person}{Suyash Fulay} {and} \bibinfo{person}{Deb Roy}.} \bibinfo{year}{2025}\natexlab{}.
\newblock \showarticletitle{The Empty Chair: Using LLMs to Raise Missing Perspectives in Policy Deliberations}.
\newblock \bibinfo{journal}{\emph{arXiv preprint arXiv:2503.13812}} (\bibinfo{year}{2025}).
\newblock


\bibitem[Gajos and Mamykina(2022)]%
        {gajosPeopleEngageCognitively2022}
\bibfield{author}{\bibinfo{person}{Krzysztof~Z. Gajos} {and} \bibinfo{person}{Lena Mamykina}.} \bibinfo{year}{2022}\natexlab{}.
\newblock \showarticletitle{Do {{People Engage Cognitively}} with {{AI}}? {{Impact}} of {{AI Assistance}} on {{Incidental Learning}}}. In \bibinfo{booktitle}{\emph{27th {{International Conference}} on {{Intelligent User Interfaces}}}}. \bibinfo{publisher}{ACM}, \bibinfo{address}{Helsinki Finland}, \bibinfo{pages}{794--806}.
\newblock
\showISBNx{978-1-4503-9144-3}
\href{https://doi.org/10.1145/3490099.3511138}{doi:\nolinkurl{10.1145/3490099.3511138}}


\bibitem[Gao et~al\mbox{.}(2023)]%
        {gaocollabcoder}
\bibfield{author}{\bibinfo{person}{Jie Gao}, \bibinfo{person}{Yuchen Guo}, \bibinfo{person}{Toby Jia-Jun Li}, {and} \bibinfo{person}{Simon~Tangi Perrault}.} \bibinfo{year}{2023}\natexlab{}.
\newblock \showarticletitle{CollabCoder: A GPT-Powered WorkFlow for Collaborative Qualitative Analysis}. In \bibinfo{booktitle}{\emph{Companion Publication of the 2023 Conference on Computer Supported Cooperative Work and Social Computing}} (Minneapolis, MN, USA) \emph{(\bibinfo{series}{CSCW '23 Companion})}. \bibinfo{publisher}{Association for Computing Machinery}, \bibinfo{address}{New York, NY, USA}, \bibinfo{pages}{354–357}.
\newblock
\showISBNx{9798400701290}
\href{https://doi.org/10.1145/3584931.3607500}{doi:\nolinkurl{10.1145/3584931.3607500}}


\bibitem[García-Marzá and Calvo(2024)]%
        {GarciaMarza2024Algorithmic}
\bibfield{author}{\bibinfo{person}{Domingo García-Marzá} {and} \bibinfo{person}{Patrici Calvo}.} \bibinfo{year}{2024}\natexlab{}.
\newblock \bibinfo{booktitle}{\emph{Algorithmic Democracy: A Critical Perspective Based on Deliberative Democracy} (\bibinfo{edition}{1} ed.)}. \bibinfo{series}{Philosophy and Politics - Critical Explorations}, Vol.~\bibinfo{volume}{29}.
\newblock \bibinfo{publisher}{Springer Cham}. XIII + 257 pages.
\newblock
\showISBNx{978-3-031-53014-2}
\href{https://doi.org/10.1007/978-3-031-53015-9}{doi:\nolinkurl{10.1007/978-3-031-53015-9}}


\bibitem[Gebreegziabher et~al\mbox{.}(2023)]%
        {gebreegziabher2023patat}
\bibfield{author}{\bibinfo{person}{Simret~Araya Gebreegziabher}, \bibinfo{person}{Zheng Zhang}, \bibinfo{person}{Xiaohang Tang}, \bibinfo{person}{Yihao Meng}, \bibinfo{person}{Elena~L Glassman}, {and} \bibinfo{person}{Toby Jia-Jun Li}.} \bibinfo{year}{2023}\natexlab{}.
\newblock \showarticletitle{Patat: Human-ai collaborative qualitative coding with explainable interactive rule synthesis}. In \bibinfo{booktitle}{\emph{Proceedings of the 2023 CHI Conference on Human Factors in Computing Systems}}. \bibinfo{pages}{1--19}.
\newblock


\bibitem[Gillani et~al\mbox{.}(2023)]%
        {gillani2023air}
\bibfield{author}{\bibinfo{person}{Nabeel Gillani}, \bibinfo{person}{Cassandra Overney}, \bibinfo{person}{Claire Schuch}, {and} \bibinfo{person}{Kumar Chandra}.} \bibinfo{year}{2023}\natexlab{}.
\newblock \bibinfo{title}{Fostering More Integrated Schools Through Community-Driven, Machine-Informed Rezoning}.
\newblock \bibinfo{howpublished}{Essay for the American Institutes of Research, Integration and Equity 2.0: New Approaches for a New Era}.
\newblock
\urldef\tempurl%
\url{https://edopportunity.org/docs/segregation/resources/AIR%20-%20Integration%20and%20Equity%202.0.pdf}
\showURL{%
\tempurl}
\newblock
\shownote{Pages 123 to 134}.


\bibitem[Glaser et~al\mbox{.}(1968)]%
        {glaser1968discovery}
\bibfield{author}{\bibinfo{person}{Barney~G Glaser}, \bibinfo{person}{Anselm~L Strauss}, {and} \bibinfo{person}{Elizabeth Strutzel}.} \bibinfo{year}{1968}\natexlab{}.
\newblock \showarticletitle{The discovery of grounded theory; strategies for qualitative research}.
\newblock \bibinfo{journal}{\emph{Nursing research}} \bibinfo{volume}{17}, \bibinfo{number}{4} (\bibinfo{year}{1968}), \bibinfo{pages}{364}.
\newblock


\bibitem[Grandeit et~al\mbox{.}(2020)]%
        {grandeit2020using}
\bibfield{author}{\bibinfo{person}{Philipp Grandeit}, \bibinfo{person}{Carolyn Haberkern}, \bibinfo{person}{Maximiliane Lang}, \bibinfo{person}{Jens Albrecht}, {and} \bibinfo{person}{Robert Lehmann}.} \bibinfo{year}{2020}\natexlab{}.
\newblock \showarticletitle{Using BERT for qualitative content analysis in psychosocial online counseling}. In \bibinfo{booktitle}{\emph{Proceedings of the Fourth Workshop on Natural Language Processing and Computational Social Science}}. \bibinfo{pages}{11--23}.
\newblock


\bibitem[Grimmer and Stewart(2013)]%
        {grimmer2013text}
\bibfield{author}{\bibinfo{person}{Justin Grimmer} {and} \bibinfo{person}{Brandon~M Stewart}.} \bibinfo{year}{2013}\natexlab{}.
\newblock \showarticletitle{Text as data: The promise and pitfalls of automatic content analysis methods for political texts}.
\newblock \bibinfo{journal}{\emph{Political analysis}} \bibinfo{volume}{21}, \bibinfo{number}{3} (\bibinfo{year}{2013}), \bibinfo{pages}{267--297}.
\newblock


\bibitem[Gün et~al\mbox{.}(2020)]%
        {gunUrbanDesignEmpowerment2020}
\bibfield{author}{\bibinfo{person}{Ahmet Gün}, \bibinfo{person}{Yüksel Demir}, {and} \bibinfo{person}{Burak Pak}.} \bibinfo{year}{2020}\natexlab{}.
\newblock \showarticletitle{Urban Design Empowerment through {{ICT-based}} Platforms in {{Europe}}}.
\newblock  \bibinfo{volume}{24}, \bibinfo{number}{2} (\bibinfo{year}{2020}), \bibinfo{pages}{189--215}.
\newblock
\showISSN{1226-5934, 2161-6779}
\href{https://doi.org/10.1080/12265934.2019.1604250}{doi:\nolinkurl{10.1080/12265934.2019.1604250}}


\bibitem[Hagmann et~al\mbox{.}(2024)]%
        {hagmann2024personal}
\bibfield{author}{\bibinfo{person}{David Hagmann}, \bibinfo{person}{Julia~A Minson}, {and} \bibinfo{person}{Catherine~H Tinsley}.} \bibinfo{year}{2024}\natexlab{}.
\newblock \showarticletitle{Personal narratives build trust across ideological divides.}
\newblock \bibinfo{journal}{\emph{Journal of Applied Psychology}} (\bibinfo{year}{2024}).
\newblock


\bibitem[Hitch(2024)]%
        {hitch2024artificial}
\bibfield{author}{\bibinfo{person}{Danielle Hitch}.} \bibinfo{year}{2024}\natexlab{}.
\newblock \showarticletitle{Artificial intelligence augmented qualitative analysis: the way of the future?}
\newblock \bibinfo{journal}{\emph{Qualitative Health Research}} \bibinfo{volume}{34}, \bibinfo{number}{7} (\bibinfo{year}{2024}), \bibinfo{pages}{595--606}.
\newblock


\bibitem[Horvath et~al\mbox{.}(2023)]%
        {horvath2023citizens}
\bibfield{author}{\bibinfo{person}{Laszlo Horvath}, \bibinfo{person}{Oliver James}, \bibinfo{person}{Susan Banducci}, {and} \bibinfo{person}{Ana Beduschi}.} \bibinfo{year}{2023}\natexlab{}.
\newblock \showarticletitle{Citizens' acceptance of artificial intelligence in public services: Evidence from a conjoint experiment about processing permit applications}.
\newblock \bibinfo{journal}{\emph{Government Information Quarterly}} \bibinfo{volume}{40}, \bibinfo{number}{4} (\bibinfo{year}{2023}), \bibinfo{pages}{101876}.
\newblock


\bibitem[Huang et~al\mbox{.}(2024)]%
        {huang2024training}
\bibfield{author}{\bibinfo{person}{Chengyu Huang}, \bibinfo{person}{Zeqiu Wu}, \bibinfo{person}{Yushi Hu}, {and} \bibinfo{person}{Wenya Wang}.} \bibinfo{year}{2024}\natexlab{}.
\newblock \showarticletitle{Training language models to generate text with citations via fine-grained rewards}.
\newblock \bibinfo{journal}{\emph{arXiv preprint arXiv:2402.04315}} (\bibinfo{year}{2024}).
\newblock


\bibitem[Huang et~al\mbox{.}(2023)]%
        {huang2023survey}
\bibfield{author}{\bibinfo{person}{Lei Huang}, \bibinfo{person}{Weijiang Yu}, \bibinfo{person}{Weitao Ma}, \bibinfo{person}{Weihong Zhong}, \bibinfo{person}{Zhangyin Feng}, \bibinfo{person}{Haotian Wang}, \bibinfo{person}{Qianglong Chen}, \bibinfo{person}{Weihua Peng}, \bibinfo{person}{Xiaocheng Feng}, \bibinfo{person}{Bing Qin}, {et~al\mbox{.}}} \bibinfo{year}{2023}\natexlab{}.
\newblock \showarticletitle{A survey on hallucination in large language models: Principles, taxonomy, challenges, and open questions}.
\newblock \bibinfo{journal}{\emph{ACM Transactions on Information Systems}} (\bibinfo{year}{2023}).
\newblock


\bibitem[Hughes et~al\mbox{.}(2025)]%
        {hughes2025voice}
\bibfield{author}{\bibinfo{person}{Maggie Hughes}, \bibinfo{person}{Cassandra Overney}, \bibinfo{person}{Ashima Kamra}, \bibinfo{person}{Jasmin Tepale}, \bibinfo{person}{Elizabeth Hamby}, \bibinfo{person}{Mahmood Jasim}, {and} \bibinfo{person}{Deb Roy}.} \bibinfo{year}{2025}\natexlab{}.
\newblock \showarticletitle{Voice to Vision: Enhancing Civic Decision-Making through Co-Designed Data Infrastructure}.
\newblock \bibinfo{journal}{\emph{arXiv preprint arXiv:2505.14853}} (\bibinfo{year}{2025}).
\newblock


\bibitem[Hullman and Diakopoulos(2011)]%
        {hullman2011visualization}
\bibfield{author}{\bibinfo{person}{Jessica Hullman} {and} \bibinfo{person}{Nick Diakopoulos}.} \bibinfo{year}{2011}\natexlab{}.
\newblock \showarticletitle{Visualization rhetoric: Framing effects in narrative visualization}.
\newblock \bibinfo{journal}{\emph{IEEE transactions on visualization and computer graphics}} \bibinfo{volume}{17}, \bibinfo{number}{12} (\bibinfo{year}{2011}), \bibinfo{pages}{2231--2240}.
\newblock


\bibitem[Irvin and Stansbury(2004)]%
        {irvinCitizenParticipationDecision2004}
\bibfield{author}{\bibinfo{person}{Ren{\'e}e~A. Irvin} {and} \bibinfo{person}{John Stansbury}.} \bibinfo{year}{2004}\natexlab{}.
\newblock \showarticletitle{Citizen {{Participation}} in {{Decision Making}}: {{Is It Worth}} the {{Effort}}?}
\newblock \bibinfo{journal}{\emph{Public Administration Review}} \bibinfo{volume}{64}, \bibinfo{number}{1} (\bibinfo{year}{2004}), \bibinfo{pages}{55--65}.
\newblock
\showISSN{1540-6210}
\href{https://doi.org/10.1111/j.1540-6210.2004.00346.x}{doi:\nolinkurl{10.1111/j.1540-6210.2004.00346.x}}


\bibitem[Jasim et~al\mbox{.}(2021a)]%
        {jasimCommunityPulseFacilitatingCommunity2021}
\bibfield{author}{\bibinfo{person}{Mahmood Jasim}, \bibinfo{person}{Enamul Hoque}, \bibinfo{person}{Ali Sarvghad}, {and} \bibinfo{person}{Narges Mahyar}.} \bibinfo{year}{2021}\natexlab{a}.
\newblock \showarticletitle{{{CommunityPulse}}: {{Facilitating Community Input Analysis}} by {{Surfacing Hidden Insights}}, {{Reflections}}, and {{Priorities}}}. In \bibinfo{booktitle}{\emph{Designing {{Interactive Systems Conference}} 2021}}. \bibinfo{publisher}{ACM}, \bibinfo{address}{Virtual Event USA}, \bibinfo{pages}{846--863}.
\newblock
\showISBNx{978-1-4503-8476-6}
\href{https://doi.org/10.1145/3461778.3462132}{doi:\nolinkurl{10.1145/3461778.3462132}}


\bibitem[Jasim et~al\mbox{.}(2021b)]%
        {jasimCommunityClickCapturingReporting2021}
\bibfield{author}{\bibinfo{person}{Mahmood Jasim}, \bibinfo{person}{Pooya Khaloo}, \bibinfo{person}{Somin Wadhwa}, \bibinfo{person}{Amy~X. Zhang}, \bibinfo{person}{Ali Sarvghad}, {and} \bibinfo{person}{Narges Mahyar}.} \bibinfo{year}{2021}\natexlab{b}.
\newblock \showarticletitle{{{CommunityClick}}: {{Capturing}} and {{Reporting Community Feedback}} from {{Town Halls}} to {{Improve Inclusivity}}}.
\newblock \bibinfo{journal}{\emph{Proceedings of the ACM on Human-Computer Interaction}} \bibinfo{volume}{4}, \bibinfo{number}{CSCW3} (\bibinfo{date}{Jan.} \bibinfo{year}{2021}), \bibinfo{pages}{213:1--213:32}.
\newblock
\href{https://doi.org/10.1145/3432912}{doi:\nolinkurl{10.1145/3432912}}


\bibitem[Jiang et~al\mbox{.}(2024a)]%
        {jiang2024bridging}
\bibfield{author}{\bibinfo{person}{Hang Jiang}, \bibinfo{person}{Doug Beeferman}, \bibinfo{person}{William Brannon}, \bibinfo{person}{Andrew Heyward}, {and} \bibinfo{person}{Deb Roy}.} \bibinfo{year}{2024}\natexlab{a}.
\newblock \showarticletitle{Bridging Dictionary: AI-Generated Dictionary of Partisan Language Use}. In \bibinfo{booktitle}{\emph{Companion Publication of the 2024 Conference on Computer-Supported Cooperative Work and Social Computing}}. \bibinfo{pages}{79--82}.
\newblock


\bibitem[Jiang et~al\mbox{.}(2024b)]%
        {jiang-etal-2024-personallm}
\bibfield{author}{\bibinfo{person}{Hang Jiang}, \bibinfo{person}{Xiajie Zhang}, \bibinfo{person}{Xubo Cao}, \bibinfo{person}{Cynthia Breazeal}, \bibinfo{person}{Deb Roy}, {and} \bibinfo{person}{Jad Kabbara}.} \bibinfo{year}{2024}\natexlab{b}.
\newblock \showarticletitle{PersonaLLM: Investigating the Ability of Large Language Models to Express Personality Traits}. In \bibinfo{booktitle}{\emph{Findings of the Association for Computational Linguistics: NAACL 2024}}. \bibinfo{publisher}{Association for Computational Linguistics}, \bibinfo{pages}{3605--3627}.
\newblock


\bibitem[Jiang et~al\mbox{.}(2024c)]%
        {jiang-etal-2024-leveraging}
\bibfield{author}{\bibinfo{person}{Hang Jiang}, \bibinfo{person}{Xiajie Zhang}, \bibinfo{person}{Robert Mahari}, \bibinfo{person}{Daniel Kessler}, \bibinfo{person}{Eric Ma}, \bibinfo{person}{Tal August}, \bibinfo{person}{Irene Li}, \bibinfo{person}{Alex Pentland}, \bibinfo{person}{Yoon Kim}, \bibinfo{person}{Deb Roy}, {and} \bibinfo{person}{Jad Kabbara}.} \bibinfo{year}{2024}\natexlab{c}.
\newblock \showarticletitle{Leveraging Large Language Models for Learning Complex Legal Concepts through Storytelling}. In \bibinfo{booktitle}{\emph{Proceedings of the 62nd Annual Meeting of the Association for Computational Linguistics (Volume 1: Long Papers)}}. \bibinfo{publisher}{Association for Computational Linguistics}, \bibinfo{pages}{7194--7219}.
\newblock


\bibitem[Jiang et~al\mbox{.}(2021)]%
        {jiang2021supporting}
\bibfield{author}{\bibinfo{person}{Jialun~Aaron Jiang}, \bibinfo{person}{Kandrea Wade}, \bibinfo{person}{Casey Fiesler}, {and} \bibinfo{person}{Jed~R Brubaker}.} \bibinfo{year}{2021}\natexlab{}.
\newblock \showarticletitle{Supporting serendipity: Opportunities and challenges for Human-AI Collaboration in qualitative analysis}.
\newblock \bibinfo{journal}{\emph{Proceedings of the ACM on Human-Computer Interaction}} \bibinfo{volume}{5}, \bibinfo{number}{CSCW1} (\bibinfo{year}{2021}), \bibinfo{pages}{1--23}.
\newblock


\bibitem[Johnson et~al\mbox{.}(2017)]%
        {johnsonCommunityConversationalSupporting2017}
\bibfield{author}{\bibinfo{person}{Ian~G. Johnson}, \bibinfo{person}{Alistair MacDonald}, \bibinfo{person}{Jo Briggs}, \bibinfo{person}{Jennifer Manuel}, \bibinfo{person}{Karen Salt}, \bibinfo{person}{Emma Flynn}, {and} \bibinfo{person}{John Vines}.} \bibinfo{year}{2017}\natexlab{}.
\newblock \showarticletitle{Community {{Conversational}}: {{Supporting}} and {{Capturing Deliberative Talk}} in {{Local Consultation Processes}}}. In \bibinfo{booktitle}{\emph{Proceedings of the 2017 {{CHI Conference}} on {{Human Factors}} in {{Computing Systems}}}} \emph{(\bibinfo{series}{{{CHI}} '17})}. \bibinfo{publisher}{Association for Computing Machinery}, \bibinfo{address}{New York, NY, USA}, \bibinfo{pages}{2320--2333}.
\newblock
\showISBNx{978-1-4503-4655-9}
\href{https://doi.org/10.1145/3025453.3025559}{doi:\nolinkurl{10.1145/3025453.3025559}}


\bibitem[Kabbara et~al\mbox{.}(2025)]%
        {kabbara2025ai}
\bibfield{author}{\bibinfo{person}{Jad Kabbara}, \bibinfo{person}{Thanh-Mai Phan}, \bibinfo{person}{Marina Rakhilin}, \bibinfo{person}{Maya~E Detwiller}, \bibinfo{person}{Dimitra Dimitrakopoulou}, {and} \bibinfo{person}{Deb Roy}.} \bibinfo{year}{2025}\natexlab{}.
\newblock \showarticletitle{AI-assisted sensemaking: Human-AI collaboration for the analysis and interpretation of recorded facilitated conversations}. In \bibinfo{booktitle}{\emph{Proceedings of the Extended Abstracts of the CHI Conference on Human Factors in Computing Systems}}. \bibinfo{pages}{1--8}.
\newblock


\bibitem[Kalla and Broockman(2023)]%
        {kallaWhichNarrativeStrategies2023}
\bibfield{author}{\bibinfo{person}{Joshua~L. Kalla} {and} \bibinfo{person}{David~E. Broockman}.} \bibinfo{year}{2023}\natexlab{}.
\newblock \showarticletitle{Which {{Narrative Strategies Durably Reduce Prejudice}}? {{Evidence}} from {{Field}} and {{Survey Experiments Supporting}} the {{Efficacy}} of {{Perspective-Getting}}}.
\newblock \bibinfo{journal}{\emph{American Journal of Political Science}} \bibinfo{volume}{67}, \bibinfo{number}{1} (\bibinfo{year}{2023}), \bibinfo{pages}{185--204}.
\newblock
\showISSN{1540-5907}
\href{https://doi.org/10.1111/ajps.12657}{doi:\nolinkurl{10.1111/ajps.12657}}


\bibitem[Khattab et~al\mbox{.}(2023)]%
        {khattab2023dspy}
\bibfield{author}{\bibinfo{person}{Omar Khattab}, \bibinfo{person}{Arnav Singhvi}, \bibinfo{person}{Paridhi Maheshwari}, \bibinfo{person}{Zhiyuan Zhang}, \bibinfo{person}{Keshav Santhanam}, \bibinfo{person}{Sri Vardhamanan}, \bibinfo{person}{Saiful Haq}, \bibinfo{person}{Ashutosh Sharma}, \bibinfo{person}{Thomas~T. Joshi}, \bibinfo{person}{Hanna Moazam}, \bibinfo{person}{Heather Miller}, \bibinfo{person}{Matei Zaharia}, {and} \bibinfo{person}{Christopher Potts}.} \bibinfo{year}{2023}\natexlab{}.
\newblock \showarticletitle{DSPy: Compiling Declarative Language Model Calls into Self-Improving Pipelines}.
\newblock \bibinfo{journal}{\emph{arXiv preprint arXiv:2310.03714}} (\bibinfo{year}{2023}).
\newblock


\bibitem[Kriplean et~al\mbox{.}(2012)]%
        {kripleanSupportingReflectivePublic2012}
\bibfield{author}{\bibinfo{person}{Travis Kriplean}, \bibinfo{person}{Jonathan Morgan}, \bibinfo{person}{Deen Freelon}, \bibinfo{person}{Alan Borning}, {and} \bibinfo{person}{Lance Bennett}.} \bibinfo{year}{2012}\natexlab{}.
\newblock \showarticletitle{Supporting Reflective Public Thought with Considerit}. In \bibinfo{booktitle}{\emph{Proceedings of the {{ACM}} 2012 Conference on {{Computer Supported Cooperative Work}}}} \emph{(\bibinfo{series}{{{CSCW}} '12})}. \bibinfo{publisher}{Association for Computing Machinery}, \bibinfo{address}{New York, NY, USA}, \bibinfo{pages}{265--274}.
\newblock
\showISBNx{978-1-4503-1086-4}
\href{https://doi.org/10.1145/2145204.2145249}{doi:\nolinkurl{10.1145/2145204.2145249}}


\bibitem[Kubin et~al\mbox{.}(2021)]%
        {kubinPersonalExperiencesBridge2021}
\bibfield{author}{\bibinfo{person}{Emily Kubin}, \bibinfo{person}{Curtis Puryear}, \bibinfo{person}{Chelsea Schein}, {and} \bibinfo{person}{Kurt Gray}.} \bibinfo{year}{2021}\natexlab{}.
\newblock \showarticletitle{Personal Experiences Bridge Moral and Political Divides Better than Facts}.
\newblock \bibinfo{journal}{\emph{Proceedings of the National Academy of Sciences}} \bibinfo{volume}{118}, \bibinfo{number}{6} (\bibinfo{date}{Feb.} \bibinfo{year}{2021}), \bibinfo{pages}{e2008389118}.
\newblock
\href{https://doi.org/10.1073/pnas.2008389118}{doi:\nolinkurl{10.1073/pnas.2008389118}}


\bibitem[Lambert(2013)]%
        {lambert2013digital}
\bibfield{author}{\bibinfo{person}{Joe Lambert}.} \bibinfo{year}{2013}\natexlab{}.
\newblock \bibinfo{booktitle}{\emph{Digital storytelling: Capturing lives, creating community}}.
\newblock \bibinfo{publisher}{Routledge}.
\newblock


\bibitem[Lee et~al\mbox{.}(2019)]%
        {leeWeBuildAIParticipatoryFramework2019}
\bibfield{author}{\bibinfo{person}{Min~Kyung Lee}, \bibinfo{person}{Daniel Kusbit}, \bibinfo{person}{Anson Kahng}, \bibinfo{person}{Ji~Tae Kim}, \bibinfo{person}{Xinran Yuan}, \bibinfo{person}{Allissa Chan}, \bibinfo{person}{Daniel See}, \bibinfo{person}{Ritesh Noothigattu}, \bibinfo{person}{Siheon Lee}, \bibinfo{person}{Alexandros Psomas}, {and} \bibinfo{person}{Ariel~D. Procaccia}.} \bibinfo{year}{2019}\natexlab{}.
\newblock \showarticletitle{{{WeBuildAI}}: {{Participatory Framework}} for {{Algorithmic Governance}}}.
\newblock \bibinfo{journal}{\emph{Proceedings of the ACM on Human-Computer Interaction}} \bibinfo{volume}{3}, \bibinfo{number}{CSCW} (\bibinfo{date}{Nov.} \bibinfo{year}{2019}), \bibinfo{pages}{1--35}.
\newblock
\showISSN{2573-0142}
\href{https://doi.org/10.1145/3359283}{doi:\nolinkurl{10.1145/3359283}}


\bibitem[Lennon et~al\mbox{.}(2021)]%
        {lennon2021developing}
\bibfield{author}{\bibinfo{person}{Robert~P Lennon}, \bibinfo{person}{Robbie Fraleigh}, \bibinfo{person}{Lauren~J Van~Scoy}, \bibinfo{person}{Aparna Keshaviah}, \bibinfo{person}{Xindi~C Hu}, \bibinfo{person}{Bethany~L Snyder}, \bibinfo{person}{Erin~L Miller}, \bibinfo{person}{William~A Calo}, \bibinfo{person}{Aleksandra~E Zgierska}, {and} \bibinfo{person}{Christopher Griffin}.} \bibinfo{year}{2021}\natexlab{}.
\newblock \showarticletitle{Developing and testing an automated qualitative assistant (AQUA) to support qualitative analysis}.
\newblock \bibinfo{journal}{\emph{Family Medicine and Community Health}} \bibinfo{volume}{9}, \bibinfo{number}{Suppl 1} (\bibinfo{year}{2021}).
\newblock


\bibitem[Li et~al\mbox{.}(2024)]%
        {liWhereAreWe2024}
\bibfield{author}{\bibinfo{person}{Haotian Li}, \bibinfo{person}{Yun Wang}, {and} \bibinfo{person}{Huamin Qu}.} \bibinfo{year}{2024}\natexlab{}.
\newblock \showarticletitle{Where {{Are We So Far}}? {{Understanding Data Storytelling Tools}} from the {{Perspective}} of {{Human-AI Collaboration}}}. In \bibinfo{booktitle}{\emph{Proceedings of the {{CHI Conference}} on {{Human Factors}} in {{Computing Systems}}}} \emph{(\bibinfo{series}{{{CHI}} '24})}. \bibinfo{publisher}{Association for Computing Machinery}, \bibinfo{address}{New York, NY, USA}, \bibinfo{pages}{1--19}.
\newblock
\showISBNx{9798400703300}
\href{https://doi.org/10.1145/3613904.3642726}{doi:\nolinkurl{10.1145/3613904.3642726}}


\bibitem[Mahyar et~al\mbox{.}(2016)]%
        {mahyarUDCoSpacesTableCentred2016}
\bibfield{author}{\bibinfo{person}{Narges Mahyar}, \bibinfo{person}{Kelly~J. Burke}, \bibinfo{person}{Jialiang~(Ernest) Xiang}, \bibinfo{person}{Siyi~(Cathy) Meng}, \bibinfo{person}{Kellogg~S. Booth}, \bibinfo{person}{Cynthia~L. Girling}, {and} \bibinfo{person}{Ronald~W. Kellett}.} \bibinfo{year}{2016}\natexlab{}.
\newblock \showarticletitle{{{UD Co-Spaces}}: {{A Table-Centred Multi-Display Environment}} for {{Public Engagement}} in {{Urban Design Charrettes}}}. In \bibinfo{booktitle}{\emph{Proceedings of the 2016 {{ACM International Conference}} on {{Interactive Surfaces}} and {{Spaces}}}} \emph{(\bibinfo{series}{{{ISS}} '16})}. \bibinfo{publisher}{Association for Computing Machinery}, \bibinfo{address}{New York, NY, USA}, \bibinfo{pages}{109--118}.
\newblock
\showISBNx{978-1-4503-4248-3}
\href{https://doi.org/10.1145/2992154.2992163}{doi:\nolinkurl{10.1145/2992154.2992163}}


\bibitem[Mahyar et~al\mbox{.}(2019)]%
        {mahyarCivicDataDeluge2019}
\bibfield{author}{\bibinfo{person}{Narges Mahyar}, \bibinfo{person}{Diana~V. Nguyen}, \bibinfo{person}{Maggie Chan}, \bibinfo{person}{Jiayi Zheng}, {and} \bibinfo{person}{Steven~P. Dow}.} \bibinfo{year}{2019}\natexlab{}.
\newblock \showarticletitle{The {{Civic Data Deluge}}: {{Understanding}} the {{Challenges}} of {{Analyzing Large-Scale Community Input}}}. In \bibinfo{booktitle}{\emph{Proceedings of the 2019 on {{Designing Interactive Systems Conference}}}}. \bibinfo{publisher}{ACM}, \bibinfo{address}{San Diego CA USA}, \bibinfo{pages}{1171--1181}.
\newblock
\showISBNx{978-1-4503-5850-7}
\href{https://doi.org/10.1145/3322276.3322354}{doi:\nolinkurl{10.1145/3322276.3322354}}


\bibitem[Maskell et~al\mbox{.}(2018)]%
        {maskellSpokespeopleExploringRoutes2018}
\bibfield{author}{\bibinfo{person}{Thomas Maskell}, \bibinfo{person}{Clara Crivellaro}, \bibinfo{person}{Robert Anderson}, \bibinfo{person}{Tom Nappey}, \bibinfo{person}{Vera Araújo-Soares}, {and} \bibinfo{person}{Kyle Montague}.} \bibinfo{year}{2018}\natexlab{}.
\newblock \showarticletitle{Spokespeople: {Exploring} {Routes} to {Action} through {Citizen}-{Generated} {Data}}. In \bibinfo{booktitle}{\emph{Proceedings of the 2018 {CHI} {Conference} on {Human} {Factors} in {Computing} {Systems}}} \emph{(\bibinfo{series}{{CHI} '18})}. \bibinfo{publisher}{Association for Computing Machinery}, \bibinfo{address}{New York, NY, USA}, \bibinfo{pages}{1--12}.
\newblock
\showISBNx{978-1-4503-5620-6}
\href{https://doi.org/10.1145/3173574.3173979}{doi:\nolinkurl{10.1145/3173574.3173979}}


\bibitem[May and Ross(2018)]%
        {mayDesignCivicTechnology2018}
\bibfield{author}{\bibinfo{person}{Andrew May} {and} \bibinfo{person}{Tracy Ross}.} \bibinfo{year}{2018}\natexlab{}.
\newblock \showarticletitle{The design of civic technology: factors that influence public participation and impact}.
\newblock \bibinfo{journal}{\emph{Ergonomics}} \bibinfo{volume}{61}, \bibinfo{number}{2} (\bibinfo{date}{Feb.} \bibinfo{year}{2018}), \bibinfo{pages}{214--225}.
\newblock
\showISSN{0014-0139}
\href{https://doi.org/10.1080/00140139.2017.1349939}{doi:\nolinkurl{10.1080/00140139.2017.1349939}}
\newblock
\shownote{Publisher: Taylor \& Francis \_eprint: https://doi.org/10.1080/00140139.2017.1349939}.


\bibitem[McAdams(2001)]%
        {mcadams2001psychology}
\bibfield{author}{\bibinfo{person}{Dan~P McAdams}.} \bibinfo{year}{2001}\natexlab{}.
\newblock \showarticletitle{The psychology of life stories}.
\newblock \bibinfo{journal}{\emph{Review of general psychology}} \bibinfo{volume}{5}, \bibinfo{number}{2} (\bibinfo{year}{2001}), \bibinfo{pages}{100--122}.
\newblock


\bibitem[Mildorf(2012)]%
        {mildorf2012second}
\bibfield{author}{\bibinfo{person}{Jarmila Mildorf}.} \bibinfo{year}{2012}\natexlab{}.
\newblock \showarticletitle{Second-person narration in literary and conversational storytelling}.
\newblock \bibinfo{journal}{\emph{Storyworlds: A Journal of Narrative Studies}}  \bibinfo{volume}{4} (\bibinfo{year}{2012}), \bibinfo{pages}{75--98}.
\newblock


\bibitem[Mildorf(2016)]%
        {mildorf2016reconsidering}
\bibfield{author}{\bibinfo{person}{Jarmila Mildorf}.} \bibinfo{year}{2016}\natexlab{}.
\newblock \showarticletitle{Reconsidering second-person narration and involvement}.
\newblock \bibinfo{journal}{\emph{Language and Literature}} \bibinfo{volume}{25}, \bibinfo{number}{2} (\bibinfo{year}{2016}), \bibinfo{pages}{145--158}.
\newblock


\bibitem[Morgan(2023)]%
        {morgan2023exploring}
\bibfield{author}{\bibinfo{person}{David~L Morgan}.} \bibinfo{year}{2023}\natexlab{}.
\newblock \showarticletitle{Exploring the use of artificial intelligence for qualitative data analysis: The case of ChatGPT}.
\newblock \bibinfo{journal}{\emph{International journal of qualitative methods}}  \bibinfo{volume}{22} (\bibinfo{year}{2023}), \bibinfo{pages}{16094069231211248}.
\newblock


\bibitem[Mugaanyi et~al\mbox{.}(2024)]%
        {mugaanyi2024evaluation}
\bibfield{author}{\bibinfo{person}{Joseph Mugaanyi}, \bibinfo{person}{Liuying Cai}, \bibinfo{person}{Sumei Cheng}, \bibinfo{person}{Caide Lu}, {and} \bibinfo{person}{Jing Huang}.} \bibinfo{year}{2024}\natexlab{}.
\newblock \showarticletitle{Evaluation of large language model performance and reliability for citations and references in scholarly writing: cross-disciplinary study}.
\newblock \bibinfo{journal}{\emph{Journal of Medical Internet Research}}  \bibinfo{volume}{26} (\bibinfo{year}{2024}), \bibinfo{pages}{e52935}.
\newblock


\bibitem[Nelimarkka(2019)]%
        {nelimarkka2019review}
\bibfield{author}{\bibinfo{person}{Matti Nelimarkka}.} \bibinfo{year}{2019}\natexlab{}.
\newblock \showarticletitle{A review of research on participation in democratic decision-making presented at SIGCHI conferences. Toward an improved trading zone between political science and HCI}.
\newblock \bibinfo{journal}{\emph{Proceedings of the ACM on Human-Computer Interaction}} \bibinfo{volume}{3}, \bibinfo{number}{CSCW} (\bibinfo{year}{2019}), \bibinfo{pages}{1--29}.
\newblock


\bibitem[Nelimarkka et~al\mbox{.}(2014)]%
        {nelimarkkaComparingThreeOnline2014}
\bibfield{author}{\bibinfo{person}{Matti Nelimarkka}, \bibinfo{person}{Brandie Nonnecke}, \bibinfo{person}{Sanjay Krishnan}, \bibinfo{person}{Tanja Aitumurto}, \bibinfo{person}{Daniel Catterson}, \bibinfo{person}{Camille Crittenden}, \bibinfo{person}{Chris Garland}, \bibinfo{person}{Conrad Gregory}, \bibinfo{person}{Ching-Chang~(Allen) Huang}, \bibinfo{person}{Gavin Newsom}, \bibinfo{person}{Jay Patel}, \bibinfo{person}{John Scott}, {and} \bibinfo{person}{Ken Goldberg}.} \bibinfo{year}{2014}\natexlab{}.
\newblock \bibinfo{title}{Comparing {Three} {Online} {Civic} {Engagement} {Platforms} using the {Spectrum} of {Public} {Participation}.}
\newblock
\urldef\tempurl%
\url{https://escholarship.org/uc/item/0bz755bj}
\showURL{%
\tempurl}


\bibitem[Nelson(2020)]%
        {nelson2020computational}
\bibfield{author}{\bibinfo{person}{Laura~K Nelson}.} \bibinfo{year}{2020}\natexlab{}.
\newblock \showarticletitle{Computational grounded theory: A methodological framework}.
\newblock \bibinfo{journal}{\emph{Sociological methods \& research}} \bibinfo{volume}{49}, \bibinfo{number}{1} (\bibinfo{year}{2020}), \bibinfo{pages}{3--42}.
\newblock


\bibitem[Overney(2025)]%
        {overney2025designing}
\bibfield{author}{\bibinfo{person}{Cassandra Overney}.} \bibinfo{year}{2025}\natexlab{}.
\newblock \showarticletitle{Designing for Constructive Civic Communication: A Framework for Human-AI Collaboration in Community Engagement Processes}.
\newblock \bibinfo{journal}{\emph{arXiv preprint arXiv:2505.11684}} (\bibinfo{year}{2025}).
\newblock


\bibitem[Overney et~al\mbox{.}(2025a)]%
        {overney2025coalesce}
\bibfield{author}{\bibinfo{person}{Cassandra Overney}, \bibinfo{person}{Daniel~T. Kessler}, \bibinfo{person}{Suyash~P. Fulay}, \bibinfo{person}{Mahmood Jasim}, {and} \bibinfo{person}{Deb Roy}.} \bibinfo{year}{2025}\natexlab{a}.
\newblock \showarticletitle{Coalesce: An Accessible Mixed-Initiative System for Designing Community-Centric Questionnaires}. In \bibinfo{booktitle}{\emph{Proceedings of the 30th International Conference on Intelligent User Interfaces (IUI '25)}} (Cagliari, Italy). \bibinfo{publisher}{ACM}, \bibinfo{address}{New York, NY, USA}, \bibinfo{pages}{24}.
\newblock
\href{https://doi.org/10.1145/3708359.3712118}{doi:\nolinkurl{10.1145/3708359.3712118}}


\bibitem[Overney et~al\mbox{.}(2025b)]%
        {overney2025boundarease}
\bibfield{author}{\bibinfo{person}{Cassandra Overney}, \bibinfo{person}{Cassandra Moe}, \bibinfo{person}{Alvin Chang}, {and} \bibinfo{person}{Nabeel Gillani}.} \bibinfo{year}{2025}\natexlab{b}.
\newblock \showarticletitle{BoundarEase: Fostering Constructive Community Engagement to Inform More Equitable Student Assignment Policies}.
\newblock \bibinfo{journal}{\emph{Proceedings of the ACM on Human-Computer Interaction}} \bibinfo{volume}{9}, \bibinfo{number}{2} (\bibinfo{date}{April} \bibinfo{year}{2025}), \bibinfo{pages}{37}.
\newblock
\href{https://doi.org/10.1145/3710938}{doi:\nolinkurl{10.1145/3710938}}


\bibitem[Overney et~al\mbox{.}(2024)]%
        {overneySenseMateAccessibleBeginnerFriendly2024}
\bibfield{author}{\bibinfo{person}{Cassandra Overney}, \bibinfo{person}{Bel{\'e}n Sald{\'i}as}, \bibinfo{person}{Dimitra Dimitrakopoulou}, {and} \bibinfo{person}{Deb Roy}.} \bibinfo{year}{2024}\natexlab{}.
\newblock \showarticletitle{{{SenseMate}}: {{An Accessible}} and {{Beginner-Friendly Human-AI Platform}} for {{Qualitative Data Analysis}}}. In \bibinfo{booktitle}{\emph{Proceedings of the 29th {{International Conference}} on {{Intelligent User Interfaces}}}} \emph{(\bibinfo{series}{{{IUI}} '24})}. \bibinfo{publisher}{Association for Computing Machinery}, \bibinfo{address}{New York, NY, USA}, \bibinfo{pages}{922--939}.
\newblock
\showISBNx{9798400705083}
\href{https://doi.org/10.1145/3640543.3645194}{doi:\nolinkurl{10.1145/3640543.3645194}}


\bibitem[Peng et~al\mbox{.}(2023)]%
        {pengPathwayUrbanPlanning2023}
\bibfield{author}{\bibinfo{person}{Zhong-Ren Peng}, \bibinfo{person}{Kai-Fa Lu}, \bibinfo{person}{Yanghe Liu}, {and} \bibinfo{person}{Wei Zhai}.} \bibinfo{year}{2023}\natexlab{}.
\newblock \showarticletitle{The {{Pathway}} of {{Urban Planning AI}}: {{From Planning Support}} to {{Plan-Making}}}.
\newblock \bibinfo{journal}{\emph{Journal of Planning Education and Research}} (\bibinfo{date}{June} \bibinfo{year}{2023}), \bibinfo{pages}{0739456X231180568}.
\newblock
\showISSN{0739-456X}
\href{https://doi.org/10.1177/0739456X231180568}{doi:\nolinkurl{10.1177/0739456X231180568}}


\bibitem[Qian et~al\mbox{.}(2024)]%
        {qian2024capacity}
\bibfield{author}{\bibinfo{person}{Haosheng Qian}, \bibinfo{person}{Yixing Fan}, \bibinfo{person}{Ruqing Zhang}, {and} \bibinfo{person}{Jiafeng Guo}.} \bibinfo{year}{2024}\natexlab{}.
\newblock \showarticletitle{On the capacity of citation generation by large language models}. In \bibinfo{booktitle}{\emph{China Conference on Information Retrieval}}. Springer, \bibinfo{pages}{109--123}.
\newblock


\bibitem[Reynante et~al\mbox{.}(2021)]%
        {reynanteFrameworkOpenCivic2021}
\bibfield{author}{\bibinfo{person}{Brandon Reynante}, \bibinfo{person}{Steven~P. Dow}, {and} \bibinfo{person}{Narges Mahyar}.} \bibinfo{year}{2021}\natexlab{}.
\newblock \showarticletitle{A {{Framework}} for {{Open Civic Design}}: {{Integrating Public Participation}}, {{Crowdsourcing}}, and {{Design Thinking}}}.
\newblock \bibinfo{journal}{\emph{Digital Government: Research and Practice}} \bibinfo{volume}{2}, \bibinfo{number}{4} (\bibinfo{date}{Oct.} \bibinfo{year}{2021}), \bibinfo{pages}{1--22}.
\newblock
\showISSN{2691-199X, 2639-0175}
\href{https://doi.org/10.1145/3487607}{doi:\nolinkurl{10.1145/3487607}}


\bibitem[Rieder and R{\"o}hle(2012)]%
        {rieder2012digital}
\bibfield{author}{\bibinfo{person}{Bernhard Rieder} {and} \bibinfo{person}{Theo R{\"o}hle}.} \bibinfo{year}{2012}\natexlab{}.
\newblock \showarticletitle{Digital methods: Five challenges}.
\newblock \bibinfo{journal}{\emph{Understanding digital humanities}} (\bibinfo{year}{2012}), \bibinfo{pages}{67--84}.
\newblock


\bibitem[Saldivar et~al\mbox{.}(2019)]%
        {saldivarCivicTechnologySocial2019}
\bibfield{author}{\bibinfo{person}{Jorge Saldivar}, \bibinfo{person}{Cristhian Parra}, \bibinfo{person}{Marcelo Alcaraz}, \bibinfo{person}{Rebeca Arteta}, {and} \bibinfo{person}{Luca Cernuzzi}.} \bibinfo{year}{2019}\natexlab{}.
\newblock \showarticletitle{Civic {{Technology}} for {{Social Innovation}}}.
\newblock \bibinfo{journal}{\emph{Computer Supported Cooperative Work (CSCW)}} \bibinfo{volume}{28}, \bibinfo{number}{1} (\bibinfo{date}{April} \bibinfo{year}{2019}), \bibinfo{pages}{169--207}.
\newblock
\showISSN{1573-7551}
\href{https://doi.org/10.1007/s10606-018-9311-7}{doi:\nolinkurl{10.1007/s10606-018-9311-7}}


\bibitem[Santurkar et~al\mbox{.}(2023)]%
        {whose_opinions}
\bibfield{author}{\bibinfo{person}{Shibani Santurkar}, \bibinfo{person}{Esin Durmus}, \bibinfo{person}{Faisal Ladhak}, \bibinfo{person}{Cinoo Lee}, \bibinfo{person}{Percy Liang}, {and} \bibinfo{person}{Tatsunori Hashimoto}.} \bibinfo{year}{2023}\natexlab{}.
\newblock \showarticletitle{Whose opinions do language models reflect?}. In \bibinfo{booktitle}{\emph{Proceedings of the 40th International Conference on Machine Learning}} (Honolulu, Hawaii, USA) \emph{(\bibinfo{series}{ICML'23})}. \bibinfo{publisher}{JMLR.org}, Article \bibinfo{articleno}{1244}, \bibinfo{numpages}{34}~pages.
\newblock


\bibitem[Sarangapani et~al\mbox{.}(2016)]%
        {sarangapani_virtualculturalcollaboration_2016}
\bibfield{author}{\bibinfo{person}{Vidya Sarangapani}, \bibinfo{person}{Ahmed Kharrufa}, \bibinfo{person}{Madeline Balaam}, \bibinfo{person}{David Leat}, {and} \bibinfo{person}{Pete Wright}.} \bibinfo{year}{2016}\natexlab{}.
\newblock \showarticletitle{Virtual.{Cultural}.{Collaboration}: mobile phones, video technology, and cross-cultural learning}. In \bibinfo{booktitle}{\emph{Proceedings of the 18th {International} {Conference} on {Human}-{Computer} {Interaction} with {Mobile} {Devices} and {Services}}} \emph{(\bibinfo{series}{{MobileHCI} '16})}. \bibinfo{publisher}{Association for Computing Machinery}, \bibinfo{address}{New York, NY, USA}, \bibinfo{pages}{341--352}.
\newblock
\showISBNx{978-1-4503-4408-1}
\href{https://doi.org/10.1145/2935334.2935354}{doi:\nolinkurl{10.1145/2935334.2935354}}


\bibitem[Schreieder et~al\mbox{.}(2025)]%
        {schreieder2025attribution}
\bibfield{author}{\bibinfo{person}{Tobias Schreieder}, \bibinfo{person}{Tim Schopf}, {and} \bibinfo{person}{Michael F{\"a}rber}.} \bibinfo{year}{2025}\natexlab{}.
\newblock \showarticletitle{Attribution, Citation, and Quotation: A Survey of Evidence-based Text Generation with Large Language Models}.
\newblock \bibinfo{journal}{\emph{arXiv preprint arXiv:2508.15396}} (\bibinfo{year}{2025}).
\newblock


\bibitem[Schroeder et~al\mbox{.}(2025)]%
        {schroeder2025forage}
\bibfield{author}{\bibinfo{person}{Hope Schroeder}, \bibinfo{person}{Doug Beeferman}, \bibinfo{person}{Maya Detwiller}, \bibinfo{person}{Dimitra Dimitrakopoulou}, {and} \bibinfo{person}{Deb Roy}.} \bibinfo{year}{2025}\natexlab{}.
\newblock \showarticletitle{Forage: Understanding RAG-based Sensemaking for Community Conversations}. In \bibinfo{booktitle}{\emph{Proceedings of the Extended Abstracts of the CHI Conference on Human Factors in Computing Systems}}. \bibinfo{pages}{1--12}.
\newblock


\bibitem[Segel and Heer(2010)]%
        {segel2010narrative}
\bibfield{author}{\bibinfo{person}{Edward Segel} {and} \bibinfo{person}{Jeffrey Heer}.} \bibinfo{year}{2010}\natexlab{}.
\newblock \showarticletitle{Narrative visualization: Telling stories with data}.
\newblock \bibinfo{journal}{\emph{IEEE transactions on visualization and computer graphics}} \bibinfo{volume}{16}, \bibinfo{number}{6} (\bibinfo{year}{2010}), \bibinfo{pages}{1139--1148}.
\newblock


\bibitem[Shaffer et~al\mbox{.}(2013)]%
        {shaffer2013effects}
\bibfield{author}{\bibinfo{person}{Victoria~A Shaffer}, \bibinfo{person}{Lukas Hulsey}, {and} \bibinfo{person}{Brian~J Zikmund-Fisher}.} \bibinfo{year}{2013}\natexlab{}.
\newblock \showarticletitle{The effects of process-focused versus experience-focused narratives in a breast cancer treatment decision task}.
\newblock \bibinfo{journal}{\emph{Patient education and counseling}} \bibinfo{volume}{93}, \bibinfo{number}{2} (\bibinfo{year}{2013}), \bibinfo{pages}{255--264}.
\newblock


\bibitem[Si et~al\mbox{.}(2023)]%
        {si2023large}
\bibfield{author}{\bibinfo{person}{Chenglei Si}, \bibinfo{person}{Navita Goyal}, \bibinfo{person}{Sherry~Tongshuang Wu}, \bibinfo{person}{Chen Zhao}, \bibinfo{person}{Shi Feng}, \bibinfo{person}{Hal Daum{\'e}~Iii}, {and} \bibinfo{person}{Jordan Boyd-Graber}.} \bibinfo{year}{2023}\natexlab{}.
\newblock \showarticletitle{Large Language Models Help Humans Verify Truthfulness--Except When They Are Convincingly Wrong}.
\newblock \bibinfo{journal}{\emph{arXiv preprint arXiv:2310.12558}} (\bibinfo{year}{2023}).
\newblock


\bibitem[Slattery et~al\mbox{.}(2024)]%
        {slattery2024systematic}
\bibfield{author}{\bibinfo{person}{P. Slattery}, \bibinfo{person}{A.~K. Saeri}, \bibinfo{person}{E.~A.~C. Grundy}, \bibinfo{person}{J. Graham}, \bibinfo{person}{M. Noetel}, \bibinfo{person}{R. Uuk}, \bibinfo{person}{J. Dao}, \bibinfo{person}{S. Pour}, \bibinfo{person}{S. Casper}, {and} \bibinfo{person}{N. Thompson}.} \bibinfo{year}{2024}\natexlab{}.
\newblock \bibinfo{title}{A Systematic Evidence Review and Common Frame of Reference for the Risks from Artificial Intelligence}.
\newblock
\href{https://doi.org/10.48550/arXiv.2408.12622}{doi:\nolinkurl{10.48550/arXiv.2408.12622}}
\showeprint[arxiv]{2408.12622}


\bibitem[Small(2021)]%
        {smallPolisScalingDeliberation2021}
\bibfield{author}{\bibinfo{person}{Christopher Small}.} \bibinfo{year}{2021}\natexlab{}.
\newblock \showarticletitle{Polis: {{Scaling Deliberation}} by {{Mapping High Dimensional Opinion Spaces}}}.
\newblock \bibinfo{journal}{\emph{RECERCA. Revista de Pensament i An{\`a}lisi}} (\bibinfo{date}{July} \bibinfo{year}{2021}).
\newblock
\showISSN{2254-4135, 1130-6149}
\href{https://doi.org/10.6035/recerca.5516}{doi:\nolinkurl{10.6035/recerca.5516}}


\bibitem[Small et~al\mbox{.}(2023)]%
        {smallOpportunitiesRisksLLMs2023}
\bibfield{author}{\bibinfo{person}{Christopher~T. Small}, \bibinfo{person}{Ivan Vendrov}, \bibinfo{person}{Esin Durmus}, \bibinfo{person}{Hadjar Homaei}, \bibinfo{person}{Elizabeth Barry}, \bibinfo{person}{Julien Cornebise}, \bibinfo{person}{Ted Suzman}, \bibinfo{person}{Deep Ganguli}, {and} \bibinfo{person}{Colin Megill}.} \bibinfo{year}{2023}\natexlab{}.
\newblock \bibinfo{title}{Opportunities and {{Risks}} of {{LLMs}} for {{Scalable Deliberation}} with {{Polis}}}.
\newblock
\href{https://doi.org/10.48550/arXiv.2306.11932}{doi:\nolinkurl{10.48550/arXiv.2306.11932}}
\showeprint[arxiv]{2306.11932}~[cs]


\bibitem[Song et~al\mbox{.}(2024)]%
        {song2024rag}
\bibfield{author}{\bibinfo{person}{Juntong Song}, \bibinfo{person}{Xingguang Wang}, \bibinfo{person}{Juno Zhu}, \bibinfo{person}{Yuanhao Wu}, \bibinfo{person}{Xuxin Cheng}, \bibinfo{person}{Randy Zhong}, {and} \bibinfo{person}{Cheng Niu}.} \bibinfo{year}{2024}\natexlab{}.
\newblock \showarticletitle{RAG-HAT: A hallucination-aware tuning pipeline for LLM in retrieval-augmented generation}. In \bibinfo{booktitle}{\emph{Proceedings of the 2024 Conference on Empirical Methods in Natural Language Processing: Industry Track}}. \bibinfo{pages}{1548--1558}.
\newblock


\bibitem[Stolper et~al\mbox{.}(2018)]%
        {stolper2018data}
\bibfield{author}{\bibinfo{person}{Charles~D Stolper}, \bibinfo{person}{Bongshin Lee}, \bibinfo{person}{Nathalie~Henry Riche}, {and} \bibinfo{person}{John Stasko}.} \bibinfo{year}{2018}\natexlab{}.
\newblock \showarticletitle{Data-driven storytelling techniques: Analysis of a curated collection of visual stories}.
\newblock In \bibinfo{booktitle}{\emph{Data-driven storytelling}}. \bibinfo{publisher}{AK Peters/CRC Press}, \bibinfo{pages}{85--105}.
\newblock


\bibitem[Sun et~al\mbox{.}(2024)]%
        {sun2024redeep}
\bibfield{author}{\bibinfo{person}{Zhongxiang Sun}, \bibinfo{person}{Xiaoxue Zang}, \bibinfo{person}{Kai Zheng}, \bibinfo{person}{Yang Song}, \bibinfo{person}{Jun Xu}, \bibinfo{person}{Xiao Zhang}, \bibinfo{person}{Weijie Yu}, {and} \bibinfo{person}{Han Li}.} \bibinfo{year}{2024}\natexlab{}.
\newblock \showarticletitle{Redeep: Detecting hallucination in retrieval-augmented generation via mechanistic interpretability}.
\newblock \bibinfo{journal}{\emph{arXiv preprint arXiv:2410.11414}} (\bibinfo{year}{2024}).
\newblock


\bibitem[Talbi(2024)]%
        {talbi2024epistemic}
\bibfield{author}{\bibinfo{person}{Merel Talbi}.} \bibinfo{year}{2024}\natexlab{}.
\newblock \showarticletitle{The Epistemic Import of Narratives}.
\newblock \bibinfo{journal}{\emph{Social Epistemology}} (\bibinfo{year}{2024}), \bibinfo{pages}{1--19}.
\newblock


\bibitem[Tanenbaum(2014)]%
        {tanenbaum2014design}
\bibfield{author}{\bibinfo{person}{Theresa~Jean Tanenbaum}.} \bibinfo{year}{2014}\natexlab{}.
\newblock \showarticletitle{Design fictional interactions: why HCI should care about stories}.
\newblock \bibinfo{journal}{\emph{interactions}} \bibinfo{volume}{21}, \bibinfo{number}{5} (\bibinfo{year}{2014}), \bibinfo{pages}{22--23}.
\newblock


\bibitem[Toukola and Ahola(2022)]%
        {toukola_digital_2022}
\bibfield{author}{\bibinfo{person}{Sebastian Toukola} {and} \bibinfo{person}{Tuomas Ahola}.} \bibinfo{year}{2022}\natexlab{}.
\newblock \showarticletitle{Digital tools for stakeholder participation in urban development projects}.
\newblock \bibinfo{journal}{\emph{Project Leadership and Society}}  \bibinfo{volume}{3} (\bibinfo{date}{Dec.} \bibinfo{year}{2022}), \bibinfo{pages}{100053}.
\newblock
\showISSN{26667215}
\href{https://doi.org/10.1016/j.plas.2022.100053}{doi:\nolinkurl{10.1016/j.plas.2022.100053}}


\bibitem[Valentini et~al\mbox{.}(2023)]%
        {valentini2023automatic}
\bibfield{author}{\bibinfo{person}{Maria Valentini}, \bibinfo{person}{Jennifer Weber}, \bibinfo{person}{Jesus Salcido}, \bibinfo{person}{T{\'e}a Wright}, \bibinfo{person}{Eliana Colunga}, {and} \bibinfo{person}{Katharina Kann}.} \bibinfo{year}{2023}\natexlab{}.
\newblock \showarticletitle{On the Automatic Generation and Simplification of Children's Stories}.
\newblock \bibinfo{journal}{\emph{arXiv preprint arXiv:2310.18502}} (\bibinfo{year}{2023}).
\newblock


\bibitem[Walters and Wilder(2023)]%
        {walters2023fabrication}
\bibfield{author}{\bibinfo{person}{William~H Walters} {and} \bibinfo{person}{Esther~Isabelle Wilder}.} \bibinfo{year}{2023}\natexlab{}.
\newblock \showarticletitle{Fabrication and errors in the bibliographic citations generated by ChatGPT}.
\newblock \bibinfo{journal}{\emph{Scientific Reports}} \bibinfo{volume}{13}, \bibinfo{number}{1} (\bibinfo{year}{2023}), \bibinfo{pages}{14045}.
\newblock


\bibitem[Xie et~al\mbox{.}(2023)]%
        {xie2023next}
\bibfield{author}{\bibinfo{person}{Zhuohan Xie}, \bibinfo{person}{Trevor Cohn}, {and} \bibinfo{person}{Jey~Han Lau}.} \bibinfo{year}{2023}\natexlab{}.
\newblock \showarticletitle{The next chapter: A study of large language models in storytelling}.
\newblock \bibinfo{journal}{\emph{arXiv preprint arXiv:2301.09790}} (\bibinfo{year}{2023}).
\newblock


\bibitem[Yao et~al\mbox{.}(2019)]%
        {yao2019plan}
\bibfield{author}{\bibinfo{person}{Lili Yao}, \bibinfo{person}{Nanyun Peng}, \bibinfo{person}{Ralph Weischedel}, \bibinfo{person}{Kevin Knight}, \bibinfo{person}{Dongyan Zhao}, {and} \bibinfo{person}{Rui Yan}.} \bibinfo{year}{2019}\natexlab{}.
\newblock \showarticletitle{Plan-and-write: Towards better automatic storytelling}. In \bibinfo{booktitle}{\emph{Proceedings of the AAAI Conference on Artificial Intelligence}}, Vol.~\bibinfo{volume}{33}. \bibinfo{pages}{7378--7385}.
\newblock


\bibitem[Ye et~al\mbox{.}(2025)]%
        {ye2025scholarmate}
\bibfield{author}{\bibinfo{person}{Runlong Ye}, \bibinfo{person}{Patrick Lee}, \bibinfo{person}{Matthew Varona}, \bibinfo{person}{Oliver Huang}, {and} \bibinfo{person}{Carolina Nobre}.} \bibinfo{year}{2025}\natexlab{}.
\newblock \showarticletitle{ScholarMate: A Mixed-Initiative Tool for Qualitative Knowledge Work and Information Sensemaking}. In \bibinfo{booktitle}{\emph{Adjunct Proceedings of the 4th Annual Symposium on Human-Computer Interaction for Work}}. \bibinfo{pages}{1--7}.
\newblock


\bibitem[Zhang et~al\mbox{.}(2023)]%
        {zhang2023qualigpt}
\bibfield{author}{\bibinfo{person}{He Zhang}, \bibinfo{person}{Chuhao Wu}, \bibinfo{person}{Jingyi Xie}, \bibinfo{person}{ChanMin Kim}, {and} \bibinfo{person}{John~M Carroll}.} \bibinfo{year}{2023}\natexlab{}.
\newblock \showarticletitle{QualiGPT: GPT as an easy-to-use tool for qualitative coding}.
\newblock \bibinfo{journal}{\emph{arXiv preprint arXiv:2310.07061}} (\bibinfo{year}{2023}).
\newblock


\bibitem[Zhang et~al\mbox{.}(2020)]%
        {zhang2020effect}
\bibfield{author}{\bibinfo{person}{Yunfeng Zhang}, \bibinfo{person}{Q~Vera Liao}, {and} \bibinfo{person}{Rachel~KE Bellamy}.} \bibinfo{year}{2020}\natexlab{}.
\newblock \showarticletitle{Effect of confidence and explanation on accuracy and trust calibration in AI-assisted decision making}. In \bibinfo{booktitle}{\emph{Proceedings of the 2020 conference on fairness, accountability, and transparency}}. \bibinfo{pages}{295--305}.
\newblock


\end{thebibliography}

\appendix


\section{Additional Data Details}
\label{app:data}

The community engagement data had two primary components: (1) a district-wide online survey and (2) a series of in-person and virtual facilitated feedback sessions. Below are additional details about these components and how the 2,480 quotes were selected and organized as material for narrative summaries. A more thorough description of the data collection and analysis process will be covered in future work.

\textbf{Survey.} A Qualtrics survey was launched to the whole district community by email and SMS, and gathered about 8,400 responses between May and December 2024. It included a variety of categorical and open-ended questions related to schools and redistricting. Most categorical questions had ``other'' responses where participants could write free-form text. This paper did not use categorical data from the survey, except for stakeholder designation (parent; student; staff; parent/staff; other, with write-in description). All write-in responses were considered for inclusion in the dataset for this paper.

\textbf{Facilitated sessions.} District staff and contracted partners led a series of in-person and virtual facilitated feedback sessions between June and December 2024.
The sessions and conversations were recorded, resulting in about 170 hours of audio across 119 events and facilitated sessions, which in total reached about 3,400 attendees.
Transcripts were segmented through a combination of human and AI effort to separate feedback highlights from facilitation and filler conversation.
Each quote was tagged with available metadata, like stakeholder type; e.g., if we knew a session was specifically held for students, then we tagged all feedback quotes as student voices.
When stakeholder metadata was unavailable (N=936 quotes out of the 2,480 used for this study), we inferred stakeholder type for the quote as parent, staff, parent/staff, student, or unknown using an LLM-driven pipeline. 

After processing, the combined survey and facilitated session data were code by topic (e.g., ``Transportation'') and by type (see \Cref{tab:code_types}). This ``type'' codebook was manually created to label the major kinds of feedback we saw in the data, and refined to improve coverage and coding accuracy. The story and personal experience quotes as the source material to generate the composite stories in this study, though the rest of the data was also used during the theme generation process for dissemination to ensure full theme coverage (see \Cref{sec:human review}). In total, 2,480 quotes ($\sim$358k words) were labeled as either type:story or type:personal experience.

\begin{table}[h]
\centering
\caption{Codebook for AI-assisted coding of quote ``type''. Quotes coded as story or personal experience were used as source material for story creation.}
\label{tab:code_types}
\begin{tabular}{lp{10cm}}
\hline
\textbf{Code} & \textbf{Description} \\
\hline
type:story & Shares a story with narrative details about themselves or someone else. \\
\hline
type:personal experience & Shares an individual's personal actions, decision-making processes, or firsthand experiences. \\
\hline
type:opinion & Expresses an opinion or reflection on an issue. \\
\hline
type:suggestion & Expresses suggestions or recommendations. \\
\hline
type:complaint & Expresses dissatisfaction or problems. \\
\hline
type:question & Expresses questions, confusions, or requests for information. \\
\hline
type:praise & Expresses positive comments or gratitude. \\
\hline
type:hypothetical & Expresses what could happen or what they would do if some future change takes place. \\
\hline
\end{tabular}
\end{table}

\section{Additional Experiment Details}
\label{app:experiments}

\subsection{Dependent variables}

Participants responded to the following items on 5-point Likert scales unless otherwise noted:

\begin{itemize}
    \item \textbf{Understanding:} ``How much do you understand the perspectives within the voices that shaped the story?'' (1 = Not at All, 5 = Completely)  
    \item \textbf{Personal Connection:} ``How connected do you feel to the voices that shaped the story?'' (1 = Not at All Connected, 5 = Extremely Connected)  
    \item \textbf{Respect:} ``How much do you value the perspective about school diversity shared in the story?'' (1 = Not at All, 5 = Extremely)  
    \item \textbf{Trust:} ``How much do you trust the voices that shaped the story?'' (1 = Not at All, 5 = Completely)  
    \item \textbf{Curiosity:} ``How much would you like to speak with the people whose voices are represented in the story?'' (1 = Not at All, 5 = Extremely)  
    \item \textbf{Change in Stance:} Difference between pre- and post-story responses to four items about increased school diversity.     Stance change was calculated as the difference between pre- and post-story responses, averaged across the four items.  Participants rated each of the following questions on a 5-point scale (1 = Strong Positive Impact, 5 = Strong Negative Impact):  
    \begin{enumerate}
        \item How do you think increased diversity in schools could impact education quality? 
        \item How do you think increased diversity in schools could impact your child’s access to opportunities?  
        \item How do you think increased diversity in schools could impact your child’s belonging or sense of community?  
        \item How do you think increased diversity in schools could impact your child’s safety at school?  
    \end{enumerate}
    \item \textbf{Change in Focus of Consideration:} Difference between pre- and post-story ratings on (a) the extent to which participants considered the impact on themselves personally, and (b) the extent to which they considered the impact on the school district as a whole.  
\end{itemize}

\subsection{Additional Experiment Analysis}
\label{app:experiment supports}

\paragraph{ANOVA \& Robustness Check} The overall ANOVA tests across the four narrative conditions showed trends in the predicted direction for \textbf{understanding} ($p=.083$) and \textbf{personal connection} ($p=.068$), but these effects did not reach significance, and \textbf{curiosity} showed no measurable differences. As a robustness check, we fit linear mixed-effects models with narrative condition and story type as fixed effects and participant as a random intercept. The results confirmed our ANOVA findings: scene-dominant ($\beta=0.32, p=.028$) and mixed ($\beta=0.34, p=.020$) narratives elicited significantly higher respect than control, and scene-dominant narratives also generated higher trust ($\beta=0.35, p=.018$). Effects remained consistent after adjusting for repeated measures and story-level variation.

\paragraph{Mediation Analysis} We also tested whether interpersonal outcomes mediated the effects of narrative condition on stance and focus of consideration. Mediation analyses showed no reliable indirect effects. In other words, while scene-dominant and mixed narratives increased respect and trust, these shifts did not translate into measurable changes in stance or consideration. 

\paragraph{Correlation Analysis} To further probe relationships among outcomes, we examined correlations between dependent variables. The interpersonal measures were strongly interrelated: respect and trust were highly correlated ($r = .78, p < .001$), with connection also closely linked to both respect ($r = .69$) and trust ($r = .61$). Curiosity showed moderate associations with connection ($r = .62$) and respect ($r = .51$). These patterns indicate that while our five interpersonal outcomes were designed as distinct constructs, participants tended to experience them together. For example, those who felt more connected to story voices also tended to respect and trust them more. In other words, narrative framing shaped a coherent cluster of relational judgments rather than isolated perceptions. By contrast, correlations between interpersonal outcomes and stance-related variables were weak to negligible, consistent with the null mediation results. Instead, stance-change items were most correlated with each other. For instance, changes in expected effects of diversity on access to opportunities were tightly linked with expected changes to belonging ($r = .55$), education quality ($r = .50$), and safety ($r = .49$). Similarly, focus-of-consideration items correlated with one another (self- vs. community-impact, $r = .31$), but showed little association with interpersonal outcomes. Overall, the correlations point to two distinct clusters of outcomes: interpersonal reactions (respect, trust, connection, understanding, curiosity) and policy-oriented judgments (stance and consideration). Narratives influenced the interpersonal cluster, but these relational shifts showed little connection to stance change, suggesting that building respect may support but not directly drive policy shifts.


\subsection{Reaction to AI Authorship}
\label{app:experiment ai reactions}

In addition to the dependent variables reported in the main text, participants were asked to evaluate the stories along three dimensions:  
\begin{enumerate}
    \item ``Generally, how would you rate the quality of the writing in the stories?'' (1 = Very Low, 5 = Very High)  
    \item ``Do you think the stories are generated by AI?'' (Yes/No)  
    \item ``They were generated by AI. How does that change your feelings towards the stories?'' (open-ended response) For participants in the control condition (who read minimally edited raw excerpts), this question was instead framed as ``They were NOT generated by AI. How does that change your feelings towards the stories?''
\end{enumerate}

\paragraph{Story Quality.}  
Across conditions, participants rated the stories as moderately high in quality. Mean scores ranged from 3.62 to 3.76 on the 5-point scale (Control: $M=3.62, SD=0.70$; Experience: $M=3.76, SD=0.66$; Mixed: $M=3.66, SD=0.66$; Opinion: $M=3.65, SD=0.60$). An ANOVA showed no significant differences across conditions, $F(3,194)=0.39, p=.762$. This suggests that overall writing quality was perceived similarly, regardless of narrative composition.  

\paragraph{Perceived AI Authorship.}  
(This is also discussed in the main text.) Participants varied in whether they believed the stories were AI-generated. The percentage of participants guessing ``Yes'' was lowest in the Control condition (28.0\%), somewhat higher for Opinion (32.7\%) and Experience (38.8\%), and highest for Mixed narratives (48.0\%). These results suggest that raw community excerpts were perceived as the most authentic, while Mixed stories, perhaps due to their blend of experiential detail and thematic framing, were most likely to be flagged as AI-generated.  

\paragraph{Reactions to AI Authorship Disclosure.}  
At the end of the study, participants were informed that the stories were AI-generated, their responses ranged from indifference to strong concern\footnote{Participants were told explicitly that the stories were generated by AI. It was implied, however, that the citations within those stories still came directly from real community responses and were not AI-generated.}. Many reported that disclosure did not alter their impressions: \emph{``It doesn’t change my feelings towards the stories''} (Opinion), \emph{``I still feel the same because these kinds of stories are typical from those who agree and disagree''} (Experience), and \emph{``Regardless of how the stories were generated, my feelings towards the stories remain unchanged''} (Mixed). Several participants even expressed surprise that the stories were not human-authored, noting \emph{``Those seemed very realistic to me''} (Experience) and \emph{``It doesn’t surprise me at this point that AI has gotten good enough to deceive me''} (Mixed).

In the \textit{Experience} group, reactions were more divided. Some participants expressed skepticism and even betrayal: \emph{``I feel I have been scammed. I thought they were real people not made up stories''} and \emph{``It makes me take them less seriously''}. Others acknowledged a shift in authenticity but still found value: \emph{``While the perspectives seemed realistic and believable, they did not feel as genuine or emotionally powerful as if I were hearing directly from real parents [...] That said, I still see value in the exercise because the narratives captured viewpoints that many parents likely hold.''}

Participants in the \textit{Mixed} condition were the most likely to question authenticity, with several stating outright devaluation: \emph{``Now that I know the stories were generated by AI, I no longer value what I read. I don’t like when AI pretends to be a human and shares human experiences''} or \emph{``I feel that it’s less credible.''} Others took a more tempered stance, noting that the perspectives remained useful: \emph{``Since AI made the stories, I see them as examples instead of real parent experiences. They are still helpful to show different opinions, but I would not take them as actual voices.''}

In the \textit{Control} group, where participants were told the stories were not AI-generated, disclosure often reinforced authenticity and trust. Many simply reported no change in perception: \emph{``It did not change my feelings at all as I did not think they were''} and \emph{``No change. I thought they were genuine human responses.''} Others emphasized that confirmation of human authorship made the stories more meaningful: \emph{``It makes them more trustworthy, because they were written by parents. So they are actual experiences you can consider and relate to,''} and \emph{``Knowing the voices are genuine makes them feel more powerful, authentic, and heartfelt, deepening my respect for their lived experiences and concerns.''} One participant noted, \emph{``I feel their voices are more genuine, trustworthy, and valuable, even if I disagree.''} Overall, disclosure in the control condition validated participants’ sense of authenticity rather than undermining it.

Overall, authorship disclosure shaped reactions differently across conditions. In the Control group, where participants were told the stories were not AI-generated, disclosure reinforced authenticity and trust, with several noting the voices felt more genuine and trustworthy. In the Experience group, learning that stories were AI-assisted prompted some unease, though many still described them as useful and reflective of real concerns. The Mixed group showed the strongest skepticism: while some acknowledged the perspectives remained valid, others reported reduced credibility and trust once AI authorship was revealed. Together, these patterns mark an authenticity boundary—participants could still find value in AI-mediated narratives, but disclosure often weakened their sense of genuineness.


\end{document}